\documentclass[twocolumn]{aastex63}
\usepackage{natbib}

\usepackage{hyperref}

\graphicspath{{./}{figures/}}
\usepackage{graphics,graphicx}
\usepackage{footnote}
\usepackage{aligned-overset}

\newcommand{\uat}[2]{\href{http://astrothesaurus.org/uat/#2}{#1 (#2)}}
\newcommand{\shortname}{PJ352$-$15}
\newcommand{\longname}{PSO J352.4034$-$15.3373}

\received{December 9, 2020}
\revised{February 3, 2021}
\accepted{February 15, 2021}
\submitjournal{ApJ}

\shorttitle{{\it Chandra} Observations of PJ352$-$15}
\shortauthors{Connor et al.}
\graphicspath{{./}{figures/}}

\begin{document}

\title{Enhanced X-ray Emission from the Most Radio-Powerful Quasar in the Universe's First Billion Years}

\correspondingauthor{Thomas Connor}
\email{thomas.p.connor@jpl.nasa.gov}

\author[0000-0002-7898-7664]{Thomas Connor}
\altaffiliation{NPP Fellow}
\affiliation{Jet Propulsion Laboratory, California Institute of Technology, 4800 Oak Grove Drive, Pasadena, CA 91109, USA}

\author[0000-0002-2931-7824]{Eduardo Ba\~nados}
\affiliation{Max Planck Institute for Astronomy, K\"onigstuhl 17, 69117 Heidelberg, Germany}
\affiliation{The Observatories of the Carnegie Institution for Science, 813 Santa Barbara St., Pasadena, CA 91101, USA}

\author[0000-0003-2686-9241]{Daniel Stern}
\affiliation{Jet Propulsion Laboratory, California Institute of Technology, 4800 Oak Grove Drive, Pasadena, CA 91109, USA}

\author[0000-0001-6647-3861]{Chris Carilli}
\affiliation{National Radio Astronomy Observatory, P.O. Box O, Socorro, NM 87801, USA}

\author[0000-0002-9378-4072]{Andrew Fabian}
\affiliation{Institute of Astronomy, Madingley Road, Cambridge CB3 0HA, UK}

\author[0000-0003-3168-5922]{Emmanuel Momjian}
\affiliation{National Radio Astronomy Observatory, P.O. Box O, Socorro, NM 87801, USA}

\author[0000-0003-2349-9310]{Sof\'ia Rojas-Ruiz}
\affiliation{Max Planck Institute for Astronomy, K\"onigstuhl 17, 69117 Heidelberg, Germany}

\author[0000-0002-2662-8803]{Roberto Decarli}
\affiliation{INAF -- Osservatorio di Astrofisica e Scienza dello Spazio di Bologna, via Gobetti 93/3, I-40129, Bologna, Italy}

\author[0000-0002-6822-2254]{Emanuele Paolo Farina}
\affiliation{Max Planck Institut f\"ur Astrophysik, Karl--Schwarzschild--Stra{\ss}e 1, D-85748, Garching bei M\"unchen, Germany}

\author[0000-0002-5941-5214]{Chiara Mazzucchelli}
\affiliation{European Southern Observatory, Alonso de Cordova 3107, Vitacura, Region Metropolitana, Chile}

\author[0000-0001-5857-5622]{Hannah P. Earnshaw}
\affiliation{Cahill Center for Astronomy and Astrophysics, California Institute of Technology, Pasadena, CA 91125, USA}

\begin{abstract}

We present deep (265 ks) \textit{Chandra} X-ray observations of PSO J352.4034$-$15.3373, a quasar at $z=5.831$ that, with a radio-to-optical flux ratio of $R>1000$, is one of the radio-loudest quasars in the early universe and is the only quasar with observed extended radio jets of kpc-scale at $z \gtrsim 6$. Modeling the X-ray spectrum of the quasar with a power law, we find a best fit of $\Gamma = 1.99^{+0.29}_{-0.28}$, leading to an X-ray luminosity of $L_{2-10} = 1.26^{+0.45}_{-0.33} \times 10^{45}\ {\rm erg}\ {\rm s}^{-1}$ and an X-ray to UV brightness ratio of $\alpha_{\rm OX} = -1.36 \pm 0.11$. We identify a diffuse structure 50 kpc (${\sim}8^{\prime\prime}$) to the NW of the quasar along the jet axis that corresponds to a $3\sigma$ enhancement in the angular density of emission and can be ruled out as a background fluctuation with a probability of $P=0.9985$. While with few detected photons the spectral fit of the structure is uncertain, we find that it has a luminosity of $L_{2-10}\sim10^{44}\ {\rm erg}\ {\rm s}^{-1}$. These observations therefore potentially represent the most distant quasar jet yet seen in X-rays. We find no evidence for excess X-ray emission where the previously-reported radio jets are seen (which have an overall linear extent of $0\farcs28$), and a bright X-ray point source located along the jet axis to the SE is revealed by optical and NIR imaging to not be associated with the quasar.

\end{abstract}


\section{Introduction} \label{sec:intro}

The evolution of supermassive black holes (SMBHs) in the early universe represents a challenge for modern cosmology, requiring significant, sustained growth from primordial seeds to explain the population of observed quasars in the first billion years of the universe \citep[e.g.,][]{2020ARA&A..58...27I}. In the past decade, not only has the number of known members of this population expanded through large surveys \citep[e.g.,][]{2016ApJS..227...11B, 2016ApJ...833..222J, 2017ApJ...849...91M, 2017MNRAS.468.4702R, 2019ApJ...883..183M, 2019AJ....157..236Y, 2019ApJ...884...30W}, but individual discoveries have pushed out the extremes of mass \citep{2015Natur.518..512W, 2020ApJ...897L..14Y}, luminosity \citep{2019MNRAS.484.5142P, 2020MNRAS.497.1842M}, and redshift \citep{2011Natur.474..616M, 2018Natur.553..473B,2021ApJ...907L...1W} that must be accounted for by theoretical models. Grappling with this challenge requires not only measuring quasar accretion rates, but also identifying mechanisms being used to produce massive growth.

One of the best ways to study this evolution is through X-ray observations, where the emission is produced in the innermost regions of the active galactic nucleus \citep[AGN,][]{2016AN....337..375F} and where, at high redshifts, the observed energies are less sensitive to  intervening obscuration. X-ray observations have been effective at investigating even the most distant known quasars \citep{2018ApJ...856L..25B}, and the results of X-ray studies of the high-redshift population include detection of variability \citep{2018A&A...614A.121N} and dual AGN (\citealt{2019ApJ...887..171C}; \citealt{2019A&A...628L...6V}). Recent works by \citet{2019A&A...630A.118V} and \citet{2021ApJ...908...53W} have also constrained the evolution of accretion physics for this population of SMBHs; they note a potential steepening of the average X-ray power law emission at high redshifts ($z \gtrsim 6$), suggestive of more rapid mass gain at the earliest epochs \citep[e.g.,][]{2013MNRAS.433.2485B}. Clearly, further study of high-redshift quasars is important for deepening our understanding of early SMBH growth, particularly when these studies expand the parameter space of analyzed quasar properties.

Radio-loud quasars are an important sub-population of high-redshift quasars for understanding early SMBH growth and evolution. Radio-loud  refers to quasars with rest-frame 5 GHz flux densities significantly greater than rest-frame optical flux densities; more formally, those quasars with radio loudness parameter $R = f_{\nu} (5\ {\rm GHz}) / f_{\nu}(4400\ \text{\normalfont \AA}) \gtrsim 10$ \citep{1989AJ.....98.1195K}. Although the fraction of radio-loud quasars remains consistent with redshift (${\sim}10$ \%; \citealt{2015ApJ...804..118B}), these objects remain effectively unstudied at high redshift in X-ray wavelengths \citep[see][]{2019A&A...630A.118V}. 
As such, characterizing the AGN properties of the earliest radio-loud quasars is a crucial step in revealing quasar growth modes.

One quasar of particular interest is \longname\ (hereafter \shortname), a radio-loud quasar at $z=5.84\pm0.02$ first reported by \citet{2018ApJ...861L..14B}. At its discovery, \shortname\ was the radio-loudest quasar known at redshifts $z \gtrsim 6$ by an order of magnitude, with $R > 1000$, although a recently-discovered blazar at $z=6.10\pm0.03$ has a similar radio loudness \citep{2020A&A...635L...7B}. 
High-resolution Very Long Baseline Interferometry (VLBI) radio imaging using the Very Long Baseline Array (VLBA) revealed the presence of linear structure at the quasar's position over 1.62 kpc ($0\farcs28$), divided into three distinct components \citep{2018ApJ...861...86M}. Whether these structures originate from a radio core with a one-sided jet or instead indicate a compact symmetric object is unclear with the currently published data\footnote{Follow-up multi-frequency VLBI analysis should address this ambiguity (E. Momjian et al., in prep)}, but it is clear that in addition to being radio-loud, \shortname\ also hosts kpc-extended radio jets. Because of these properties, \shortname\ is an excellent target for X-ray analysis; indeed, as discussed below, X-rays are potentially the best mechanism for detecting extended jet structures at this redshift \citep{2014MNRAS.442L..81F}.

In this work, we present X-ray observations of \shortname\ with {\it Chandra}. We discuss our observations in Section \ref{sec:Observations} and the X-ray properties of \shortname\ in Section \ref{sec:quasar}. In Section \ref{sec:iccmb}, we introduce the concept of inverse Compton emission from the cosmic background and detail several methods we used to detect this emission. We further describe the properties of the detected extended emission in Section \ref{sec:ext_em_disc}. Finally, we contextualize these results in Section \ref{sec:discussion}. Throughout this work, we adopt a quasar redshift of $z=5.831$ based on observed [\ion{C}{2}] $\lambda 158\mu{\rm m}$ emission (S. Rojas-Ruiz et al., in prep) and a Galactic neutral hydrogen column density of $N_{\rm H}= 1.68\times 10^{20}\,\textrm{cm}^{-2}$ in the direction of \shortname\ \citep{2016A&A...594A.116H}. We use a flat cosmology with $H_0 = 70\,\textrm{km\,s}^{-1}\,\textrm{Mpc}^{-1}$, $\Omega_M = 0.3$, and $\Omega_\Lambda = 0.7$; the scale at this redshift is $5.80\ {\rm kpc}\, \textrm{arcsec}^{-1}$. All distances given are in proper distances and errors are reported at the 1$\sigma$ (68\%) confidence level unless otherwise stated.

\section{Observations and Data Reduction}\label{sec:Observations}
\begin{deluxetable}{rcrr}
\tablecaption{{\it Chandra} Observations}
\label{tab:CXOobs}
\tablewidth{0pt}
\tablehead{
\colhead{Obs ID} & \colhead{Exposure Time} & \colhead{Start Date} & \colhead{Roll Angle}\\
\colhead{} & \colhead{(ks)} & \colhead{(YYYY-mm-dd)}& \colhead{(${}^\circ$)} }
\startdata
21415 & 41.52 &  2019-08-19 & 88\\
21416 & 19.06 &  2019-09-16 & 359\\
22728 & 59.28 &  2019-08-21 & 88\\
22729 & 45.46 &  2019-08-24 & 88\\
22730 & 38.24 &  2019-08-25 & 88\\
22850 & 31.44 &  2019-09-17 & 359\\
22851 & 29.88 &  2019-09-22 & 350	
\enddata
\end{deluxetable}

We observed \shortname\ with the Advanced CCD Imaging Spectrometer \citep[ACIS;][]{2003SPIE.4851...28G} on {\it Chandra}. Observations were conducted across seven separate visits spread across five weeks, with a total exposure time of 264.88 ks. Details of the seven visits are given in Table \ref{tab:CXOobs}. In all observations, events were recorded in the Very Faint telemetry format and with the Timed Exposure mode. {\it Chandra} was positioned so that \shortname\ appeared on the back-illuminated S3 chip during our observations. 

\begin{figure*}
\begin{center}
\includegraphics{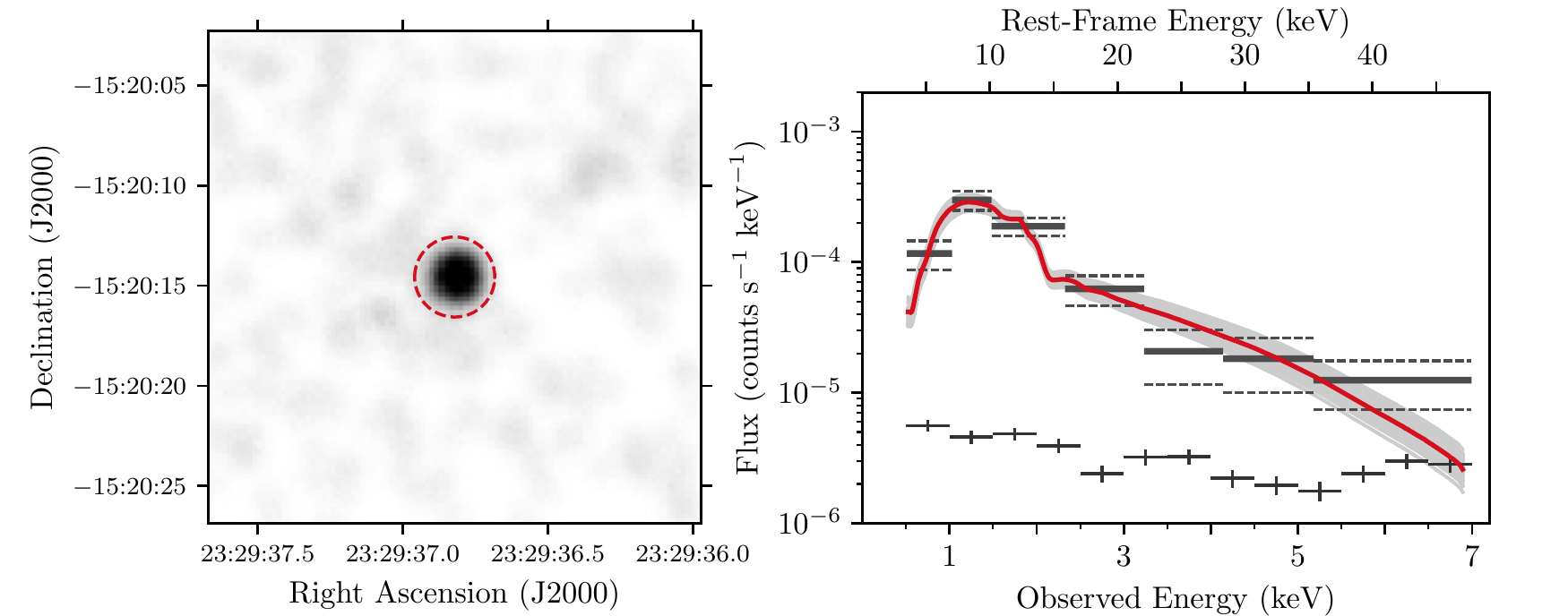}
\end{center}
\caption{{\bf Left}: $0.5-7.0$ keV {\it Chandra} observation of \shortname, smoothed by a Gaussian kernel of width $0\farcs75$. The $2\farcs0$ radius extraction region used for photometry and spectroscopy is indicated by the red circle. {\bf Right}: X-ray spectrum of \shortname. Data (dark gray) are binned for ease of display, but were not binned during fitting. The best-fitting spectrum is shown in red, while 100 spectra  with $\Delta C \leq 2.30$ from our Monte Carlo analysis are shown in gray. The background spectrum is shown in dark gray error bars at the bottom. }\label{fig:sky_and_spectrum}
\end{figure*}

Data reduction was performed using the {\it Chandra} Interactive Analysis of Observations software package \citep[CIAO,][]{2006SPIE.6270E..1VF} v4.11 with CALDB version 4.8.4.1. Reduction followed standard procedures \citep[e.g.,][]{2020ApJ...900..189C}, beginning with reprocessing using the \texttt{chandra\_repro} script with standard grade, status, and good time filters and with \texttt{VFAINT} background cleaning. As part of the standard reprocessing, events were processed with the Energy Dependent Subpixel Event Repositioning routine \citep[EDSER,][]{2004ApJ...610.1204L}; because of this reprocessing of the on-axis observations, and to enable more detailed physical modeling, all X-ray images in this work are presented at half-pixel ($0\farcs246$) resolution (and all images presented are of the combined seven observations). To allow for accurate spatial analysis by minimizing positional uncertainties, we first aligned all observations with Obs ID 22728, our deepest exposure. This alignment was done with a combination of \texttt{WAVDETECT} \citep{2002ApJS..138..185F} and the CIAO tools \texttt{wcs\_align} and \texttt{wcs\_update}. 

We used the CIAO script \texttt{merge\_obs} to generate co-added images from the seven observations. Images were generated in the soft (0.5 -- 2.0 keV), hard (2.0 -- 7.0 keV), and broad (0.5 -- 7.0 keV) bands; the image of \shortname\ in the broad band is shown in the left panel of Figure \ref{fig:sky_and_spectrum}. No significant structure is readily apparent within ${\sim}15^{\prime\prime}$ (${\sim}85$ kpc) of the quasar. For spectroscopic analysis of \shortname, we use a $2\farcs0$ circular aperture centered on the coordinates from \citet{2018ApJ...861L..14B}, which align with the X-ray centroid (Figure~\ref{fig:sky_and_spectrum}). The background was extracted from a concentric annulus with inner and outer radii of $25^{\prime\prime}$ and $38^{\prime\prime}$, respectively. Source and background spectra were created with \texttt{specextract}.

\section{X-Ray Properties of \texorpdfstring{\shortname}{P352--15}} \label{sec:quasar}
\begin{deluxetable}{rrr}
\tablecaption{X-ray Properties}
\label{tab:XProps}
\tablewidth{0pt}
\tablehead{
\colhead{Parameter} & \colhead{Value} & \colhead{Units}}
\startdata
Net Counts & $120.3^{+12.3}_{-11.2}$ & \nodata \\
Soft Counts & $80.9^{+10.1}_{-9.1}$ & \nodata \\
Hard Counts & $39.3^{+7.6}_{-6.6}$ & \nodata \\
$\mathcal{HR}$ & $-0.34^{+0.08}_{-0.09}$ & \nodata \\
$\Gamma$ & $1.99^{+0.29}_{-0.28}$ & \nodata\\
$L_{2-10} $ & $ 1.26^{+0.45}_{-0.33}\times 10^{45}$ & $\textrm{erg}\ \textrm{s}^{-1}$ \\
$F_{0.5-2.0} $ & $ 2.8^{+0.6}_{-0.5} \times 10^{-15}$ & $\textrm{erg}\ \textrm{s}^{-1}\ \textrm{cm}^{-2}$ \\
$F_{0.5-7.0} $ & $ 5.5^{+0.9}_{-0.8} \times 10^{-15}$ & $\textrm{erg}\ \textrm{s}^{-1}\ \textrm{cm}^{-2}$ \\
$L_\nu(2\ {\rm keV}) $ & $ 1.62^{+1.11}_{-0.66} \times 10^{27}$ & ${\rm erg}\ {\rm s}^{-1}\ {\rm Hz}^{-1}$\\
$L_\nu(2500\ \text{\normalfont \AA}) $ & $ 5.7^{+0.7}_{-0.6} \times 10^{30}$ & ${\rm erg}\ {\rm s}^{-1}\ {\rm Hz}^{-1}$\\
$\alpha_{\rm OX}$ & $-1.36 \pm 0.11$ & \nodata \\
\enddata
\end{deluxetable}

We detect \shortname\ in a $2\farcs0$ radius aperture with $120.3^{+12.3}_{-11.2}$ net counts in the broad band (0.5--7.0 keV), $80.9^{+10.1}_{-9.1}$ counts in the soft band (0.5--2.0 keV), and $39.3^{+7.6}_{-6.6}$ counts in the hard band (2.0--7.0 keV), with all uncertainties calculated using the method of \citet{1986ApJ...303..336G}. From these values, and using the Bayesian methodology described by \citet{2006ApJ...652..610P}, we derive a hardness ratio, $\mathcal{HR}$\footnote{$\mathcal{HR} = (H-S)/(H+S)$, where $H$ and $S$ are the net counts in the hard (2.0--7.0 keV) and soft (0.5--2.0 keV) bands, respectively.} for \shortname\ of $\mathcal{HR} = -0.34^{+0.08}_{-0.09}$. As a simple flux ratio, the hardness ratio allows for an easy comparison between sources with few detected counts, as is common for high-redshift quasars \citep{2019A&A...630A.118V}. However, for comparisons between quasars observed with different observatories or at different times, the relative differences in effective area need to be considered, and a spectral fit, when available, is thus more informative.

Spectroscopic analysis was performed using the Python-based implementation of  \texttt{XSPEC} v12.10.1 \citep{1996ASPC..101...17A}, \texttt{PyXspec}. We did not bin our spectrum, and instead used the modified C-Statistic \citep[$C$,][]{1979ApJ...228..939C, 1979ApJ...230..274W} to find the parameters of best fit. We fit the spectrum with a multiplicative combination of a power law and Galactic foreground absorption using the \texttt{XSPEC} model \texttt{phabs}$\times$\texttt{powerlaw}. Here, the Galactic absorption $N_H$ was frozen at its adopted value ($1.68\times 10^{20}\,\textrm{cm}^{-2}$), but the slope and normalization of the power law were allowed to vary. With two free parameters, $1\sigma$ uncertainties include all values with C-statistic values within $\Delta C \leq 2.30$ of the overall best fit. We used the Monte Carlo routines within \texttt{XSPEC} to evaluate the $1\sigma$ uncertainties of all parameters.
\begin{figure*}
\begin{center}
\includegraphics{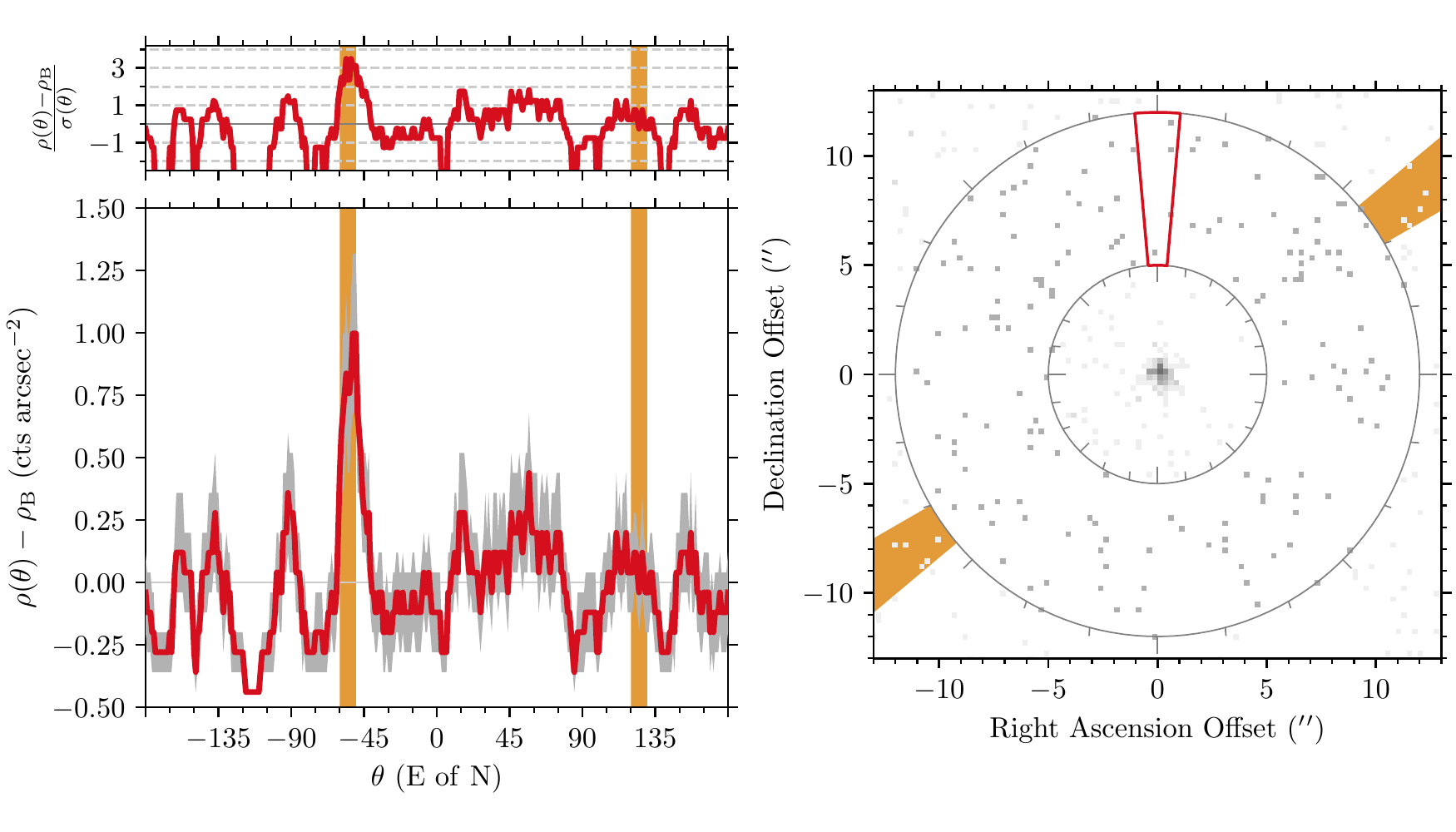}
\end{center}
\caption{{\bf Left}: angular overdensity of broad (0.5 -- 7.0 keV) emission in an annulus of radii $5\farcs0$--$12\farcs0$ centered on the quasar. Counts are summed in wedges of angular size $10^\circ$ and the contribution from the background is subtracted, as described in the text. The overdensity and its $1\sigma$ uncertainty range are marked by the red line and the gray filled region, respectively. We highlight the position angle $-55^\circ$ and its complement $125^\circ$; this is where excess emission is detected, while \citet{2018ApJ...861...86M} report jets at a position angle of $-61^\circ$. The ratio between this emission and its uncertainty is shown in the {\bf top} panel; the signal is stronger than $3\sigma$ at $-55^\circ$. {\bf Right:} broad-band sky image of \shortname, showing the extraction region (red, centered on $\theta=0^\circ$) and the $5\farcs0 - 12\farcs0$ annulus. Photons detected outside this annulus are lightened for presentation purposes. The resolved radio structure has a an overall linear extent of $0\farcs28$ (1.62 kpc), a size equivalent to one pixel in the X-ray image.}\label{fig:ang_power_spec}
\end{figure*}

From our spectral fitting, we find the emission is characterized by a power law of slope $\Gamma = 1.99^{+0.29}_{-0.28}$. This value is typical for quasars even up to $z{\sim}6$ \citep{2018A&A...614A.121N}. Including the uncertainties in the normalization of the power law, this translates to a rest-frame 2.0--10.0 keV unabsorbed luminosity of $L_{2-10} = 1.26^{+0.45}_{-0.33} \times 10^{45}\ {\rm erg}\ {\rm s}^{-1}$. This best fit is shown in Figure~\ref{fig:sky_and_spectrum}, as are 100 of the spectra explored by our Monte Carlo analysis that have $\Delta C \leq 2.30$. The X-ray properties of the quasar are summarized in Table \ref{tab:XProps}.

\vspace{4mm}

\section{Evidence for IC/CMB} \label{sec:iccmb}

\shortname\ is the source of the most distant extended (kpc-scale) radio jets yet seen (\citealt{2018ApJ...861...86M}, but see also \citealt{2020A&A...643L..12S}), and it is therefore an ideal candidate to investigate the potential for inverse Compton (IC) interactions between relativistic particles in jets and the Cosmic Microwave Background (CMB). IC/CMB has long been associated with the X-ray emission seen with jets, dating back to the first observations with {\it Chandra} \citep{2000ApJ...542..655C,2000ApJ...540L..69S}, although a number of issues have been raised against possible detections of this effect at low redshifts \citep{2015ApJ...805..154M, 2017ApJ...849...95B}.

As a simple model, IC/CMB occurs when a relativistic particle in a jet interacts with a CMB photon; the interaction depletes energy from the jets while also scattering the CMB photons to X-ray energies. The energy density of the CMB scales as $(1+z)^4$, meaning that not only is the effect of cosmological dimming countered, but at higher redshifts it becomes more likely that the CMB should dominate over magnetic fields in lobes as a mechanism for particles to radiate energy. Jets can extract rotational energy from the accretion disk \citep[e.g.,][]{1982MNRAS.199..883B}, meaning that jets can enable accretion beyond the Eddington limit, and, in the context of high redshift quasars, allow for more rapid growth \citep{2013MNRAS.432.2818G}. While the dearth of radio jets at large redshifts has been noted \citep{2015MNRAS.452.3457G}, \citet{2014MNRAS.442L..81F} proposed that if jets are primarily emitting through IC/CMB in the early universe, this could allow the growth of observed $z\gtrsim7$ SMBHs from stellar mass seeds. Direct detection of this effect is therefore of great importance to the understanding of SMBH growth.

In recent years, a number of studies have begun looking for extended X-ray emission around high-redshift AGN, but clear evidence has not yet been seen in the first billion years of the universe. \citet{2020ApJ...897..177P} reported detecting seven blazars at $3.1 \lesssim z \lesssim 4.7$ with extended emission seen by {\it Chandra}, and \citet{2020MNRAS.498.1550N} detected two extended structures on opposite angles of a $z=4.26$ radio galaxy. While \citet{2018A&A...614A.121N} identified a potential extended component associated with a $z=6.31$ quasar, further analysis by \citet{2019A&A...632A..26G} found that the reported emission is associated with a foreground structure. Finally, \citet{2014MNRAS.442L..81F} reported two potential structures around a $z=7.1$ quasar, but were unable to rule out this being caused by source confusion in the {\it XMM-Newton} observations. In this section, we describe our efforts to search for extended X-ray emission around \shortname, thereby probing potential jet-assisted growth of early quasars.

\subsection{Presence of Extended Structure} \label{ssec:ext_str}

In addition to looking for potential X-ray signatures of the radio-detected jets, we also looked for possible indications of an X-ray bright jet extending beyond the radio emission, which is roughly the size of a single ACIS pixel. Such X-ray structures have long been seen around some AGN \citep[for a review, see][]{2006ARA&A..44..463H}. While X-ray components often overlap with radio emission in AGN jets \citep[e.g.,][]{2020ApJS..250....7J}, this is not always the case; \citet{2020ApJ...904...57S}, for example, recently reported X-ray detected jet candidates around $z\sim3.2$ quasars with no corresponding radio emission. Likewise, a serendipitous discovery of a quasar X-ray jet at $z=2.5$ by \citet{2016ApJ...816L..15S} extends for ${\sim}100$ kpc, while radio emission is only seen out to $\lesssim 10$ kpc from the central AGN. In the context of IC/CMB, for a jet of fixed magnetic field, the relative flux densities between X-ray and radio emission scales as $S_X/S_r \propto (1+z)^4$ \citep{2002ARA&A..40..319C}, and so, at higher redshift, the X-ray flux will become easier to detect than the radio component. Conversely, it is also not uncommon for radio and X-ray components of jets to be spatially offset, implying their emission is generated from synchrotron emission from separate populations of jetted particles \citep[e.g.,][]{2021arXiv210102024R}; however, for such a scenario, we would expect the X-ray component to be stronger upstream in the jet relative to the radio components over large scales \citep[e.g.,][]{2007ApJ...657..145S}. 

\begin{figure*}
\begin{center}
\includegraphics{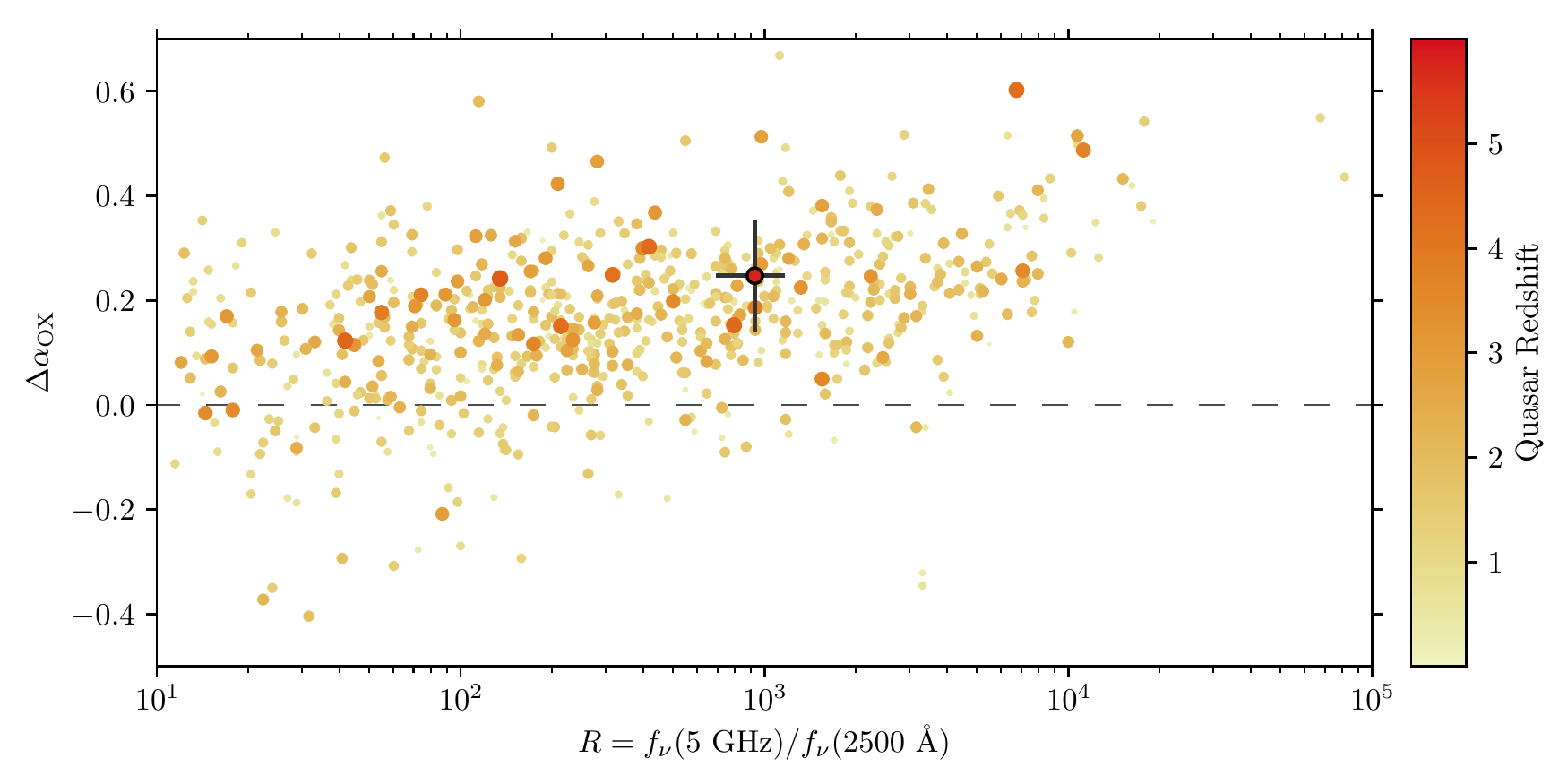}
\end{center}
\caption{Offsets from the best-fit $\alpha_{\rm OX}$ relation as a function for radio loudness for \shortname\ and a sample of radio loud quasars from \citet{2011ApJ...726...20M}. Predicted values of $\alpha_{\rm OX}$ were calculated using the nominal relation of \citet{2016ApJ...819..154L}. Points are colored by redshift, with higher-redshift quasars being redder, as indicated on the colorbar. Note that to conform to the catalog of \citet{2011ApJ...726...20M}, radio loudness in this figure is with respect to the monochromatic luminosity at 2500 \AA, not 4400 \AA. }\label{fig:delta_aox}
\end{figure*}

As a cursory inspection of the area around the quasar in X-rays (shown in Figure \ref{fig:sky_and_spectrum}) does not reveal any structure, we turn to a statistical test to search for excess emission oriented along the jet axes. If present, X-ray jets would manifest as additional X-ray emission outside of the quasar point spread function (PSF; see Section \ref{ssec:PSF_comp}) and at position angles of ${\sim}-60^\circ$ and/or ${\sim}120^\circ$, the angle and counter-angle of the jets reported by \citet{2018ApJ...861...86M}. To quantify this, we calculate $\rho(\theta)$, the azimuthal X-ray density profile \citep[e.g.,][]{2018ApJ...867...25C}, summing over all X-ray events such that
\begin{equation}
\rho(\theta) = \sum_i W_i(\theta).
\end{equation}
Here $W_i(\theta)$ is a Boolean value that evaluates to 1 if event $i$ is within an angular wedge centered on angle $\theta$ and within the wedge's inner and outer radii and to 0 otherwise. The wedge is defined by its opening angle ($\phi=10^\circ$) and inner ($r_i = 5\farcs0$) and outer ($r_o = 12\farcs0$) radii.

Although these observations were processed with EDSER to allow for event positioning to better precision than the size of an ACIS pixel, there is still an inherent uncertainty in where events were recorded. Based on the work of \citet{2003ApJ...590..586L}, we assume that all events have an inherent uncertainty in position of $\delta = 0\farcs125$ in both axes despite their repositioning. As such, we set $W_i(\theta)=1$ for events that, if shifted within $\pm \delta$ in X and/or Y positions, would fall within the wedge. The total area of the wedge is thus
\begin{equation}
A_{\rm W} = \frac{\phi \pi}{360^\circ} \left( (r_o {+} \delta)^2 - (r_i {-} \delta)^2 \right) + 2\delta\ (r_o {+} 2\delta {-} r_i).
\end{equation}
Therefore, with an expected background of surface brightness $\Omega_B$ (in units of counts ${\rm arcsec}^{-2}$), the background subtracted azimuthal density is
\begin{equation}
\rho(\theta) - \rho_B = \sum_i W_i(\theta)/ A_{\rm W} - \Omega_B.
\end{equation}
Uncertainties on this value are estimated by bootstrap resampling of all events that fall within the inner and outer radii of the annulus of interest, including the $\delta$ term. The distribution of the azimuthal X-ray emission density is shown in Figure \ref{fig:ang_power_spec}.

A clear peak is seen in Figure \ref{fig:ang_power_spec}, corresponding to a position angle of $-55^\circ$ E of N. In comparison, the jets seen by \citet{2018ApJ...861...86M} are at $-61^\circ$ E of N. From the bootstrapped uncertainties, this excess is a $3\sigma$ detection of faint structure (top panel, Figure \ref{fig:ang_power_spec}). It should be noted that these uncertainties do not include the effects of systematic choices -- the values of $r_i$, $r_o$, and $\phi$, as well as the energy range used. However, we find qualitatively similar (${\gtrsim}3\sigma$) results when using an annulus of either $3\farcs0 - 15\farcs0$ or $5\farcs0 - 10\farcs0$  or when setting $\phi \in [5^\circ,12^\circ]$. Similarly, relaxing the binary restriction on $W_i(\theta)$, by setting $W_i(\theta)=0.5$ and $W_i(\theta)=1.0$ for photons in the wedge with and without including positional uncertainties, respectively, produces no meaningful change in our results. Due to the paucity of counts, the significance of this detection is maximized in the broad energy band (0.5--7.0 keV), but the peak remains at $2-3\sigma$ significance in the soft and hard bands. 
\begin{figure*}
\begin{center}
\includegraphics{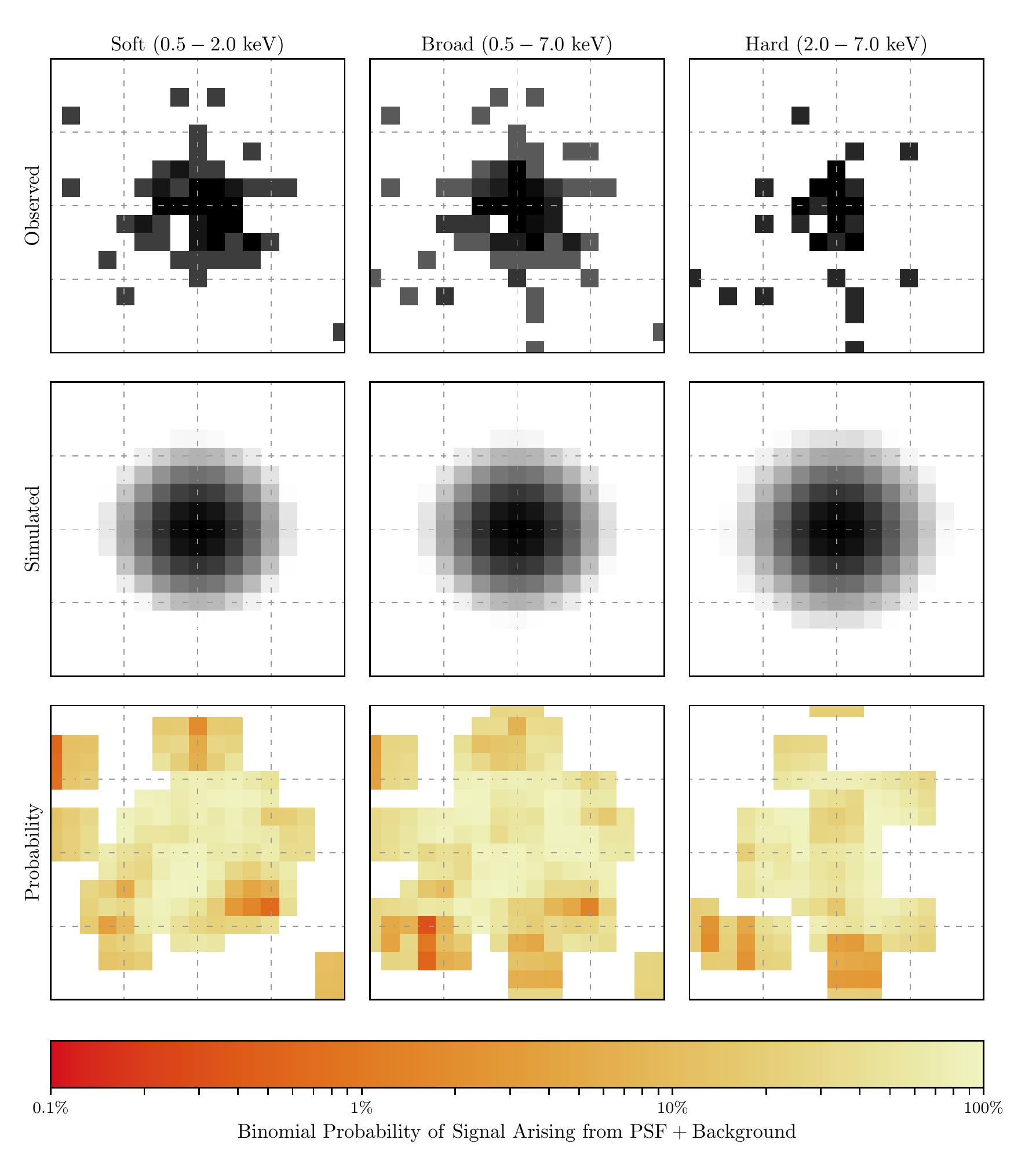}
\end{center}
\caption{Comparisons between the observed quasar ({\bf top} row) and a simulated PSF at that location ({\bf center} row), with the binomial probability of any signal arising from the combined PSF and background ({\bf bottom} row and colorbar), as described in the text. The soft ({\bf left}), broad ({\bf center}), and hard ({\bf right}) bands are independently evaluated. All images cover $4\farcs0 \times 4\farcs0$ fields-of-view centered on the peak of emission. No statistically significant deviation from the point source expectation is seen.   
}\label{fig:psf_sim}
\end{figure*}

Enhanced X-ray emission is only seen in the NW direction of the quasar; in the context of other X-ray imaging studies, the lack of emission from the counter-jet is common. In a flux-limited sample of 56 quasars with radio jets at $z\lesssim2$, \citet{2018ApJ...856...66M} report 33 have X-ray jets but none have visible counter-jets.  While tentative hints of X-ray emission from jets have been seen in quasars up to redshifts $z \lesssim 4.7$ \citep{2020ApJ...897..177P, 2020MNRAS.498.1550N}, and from jets of similar sizes ($\gtrsim50\ {\rm kpc}$) up to redshifts $z \lesssim 3.7$ \citep{2016ApJ...816L..15S, 2020MNRAS.497..988W}, this observation represents the most distant quasar with ${\gtrsim}3\sigma$ evidence for a projected X-ray jet.

\subsection{X-Ray Loudness}

We also evaluate $\alpha_{\rm OX}$, the logarithmic ratio of monochromatic luminosities between rest-frame X-ray and UV\footnote{$\alpha_{\rm OX} = 0.3838 \times \log\left(L_\nu(2\ {\rm keV}) / L_\nu(2500 \text{\normalfont \AA})\right)$, where $L_\nu$ is the monochromatic luminosity.} for \shortname. We calculate $L_\nu(2\ {\rm keV})$ directly from the best-fit power law, accounting for redshift corrections following \citet{2000AJ....119.1526S}, finding $L_\nu(2\ {\rm keV}) = 1.62^{+1.11}_{-0.66} \times 10^{27}\ {\rm erg}\ {\rm s}^{-1}\ {\rm Hz}^{-1}$. To calculate $L_\nu(2500\ \text{\normalfont \AA})$, we use the AB 1450 \AA\ absolute magnitude given by \citet{2018ApJ...861L..14B}, $M_{1450} = -25.59 \pm 0.13$ and scale that to 2500 \AA\ assuming $f_\nu \propto \nu^{\alpha_\nu}$, where, per \citet{2018ApJ...861L..14B}, we adopt $\alpha_\nu = -0.5$. From this, we calculate $L_\nu(2500\ \text{\normalfont \AA}) = 5.7^{+0.7}_{-0.6} \times 10^{30}\ {\rm erg}\ {\rm s}^{-1}\ {\rm Hz}^{-1}$ and a corresponding value of $\alpha_{\rm OX} = -1.36 \pm 0.11$.

A number of observational studies have shown a relationship between $L_\nu(2500\ \text{\normalfont \AA})$ and $\alpha_{\rm OX}$, such that, for increasing ultraviolet monochromatic luminosity, the relative strength of the X-ray luminosity declines \citep[e.g.,][]{2003AJ....125..433V, 2005AJ....130..387S, 2006AJ....131.2826S}. Using the best-fit scaling relation for radio-quiet quasars of \citet{2016ApJ...819..154L}, we would expect a value of $\alpha_{\rm OX} = -1.61$ given the ultraviolet luminosity of \shortname. Our measured value is offset from this prediction by $\Delta \alpha_{\rm OX} = 0.25 \pm 0.11$. For this offset, the uncertainty accounts for changes in both the predicted and calculated $\alpha_{\rm OX}$ with changing $L_\nu(2500\ \text{\normalfont \AA})$, but it does not include the uncertainties and intrinsic scatter in the best fit of \citet{2016ApJ...819..154L}. We show in Figure \ref{fig:delta_aox} how this value of $\Delta \alpha_{\rm OX}$ compares to a broad sample of radio-loud quasars collected by \citet{2011ApJ...726...20M}; note that, as the values of $R$ given in that work are based on $f_{\nu}(2500\ \text{\normalfont \AA})$ and not $f_{\nu}(4400\ \text{\normalfont \AA})$, we adjust the value of $R$ for \shortname\ for this plot, again extrapolating from $M_{1450}$ assuming $\alpha_\nu = -0.5$.

As can be seen in Figure \ref{fig:delta_aox}, \shortname\ is not particularly X-ray overluminous in the context of radio-loud quasars. For further comparison, we consider PSO J0309+27, the $z=6.1$ blazar \citep{2020A&A...635L...7B}. That quasar has a similar UV luminosity to \shortname\ ($M_{1450} = -25.1$) but is less X-ray overluminous ($\alpha_{\rm OX}\sim-1.6$, converting from the 10 keV-based $\tilde{\alpha}_{\rm OX}$ following \citealt{2019MNRAS.489.2732I}). \citet{2011ApJ...726...20M} argued that the lack of redshift evolution seen in $\Delta \alpha_{\rm OX}$ for radio-loud quasars was indicative that IC/CMB only produced, at most, a minor contribution to the X-ray emission from these quasars. As it stands, neither of the two radio-loudest quasars known in the early universe show enough X-ray emission to contrast with that argument.

\subsection{Point Source or Extended Emission} \label{ssec:PSF_comp}

If jets are producing excess X-ray emission around \shortname, this could potentially manifest as a deviation from the predicted {\it Chandra} PSF, assuming the jets are sufficiently separated from the AGN. To test this possibility, we compare the observed quasar in three bands -- soft (0.5 -- 2.0 keV), broad (0.5 -- 7.0 keV), and hard (2.0 -- 7.0 keV) -- to a simulated PSF for these observations. Any significant deviations, as characterized by the binomial probability of structure in the observations being caused by random samplings of the PSF and background, would be indicative of X-ray emission outside of the central point source.

For all seven observations given in Table \ref{tab:CXOobs}, we used {\tt MARX} \citep{2012SPIE.8443E..1AD} to simulate our observed point source. The input source spectrum was derived from our best-fit parameters, but with a normalization 100 times larger to produce more photons. We did not include the effects of the read-out streak or of pile-up. Each observation was simulated 100 times, so that, in total, we produced ${\sim}10,000$ simulated photons for every observed photon. We adopt an aspect blur of $0\farcs28$ to account for uncertainties in the aspect solution, following the recommendations of the {\it Chandra} X-ray Center for ACIS-S observations\footnote{\url{https://cxc.cfa.harvard.edu/ciao/why/aspectblur.html}}. The simulations were facilitated with the {\tt CIAO} task {\tt simulate\_psf}. Output events files were processed with the EDSER algorithm to be consistent with our observations. The observed quasar and simulated PSFs are shown in Figure \ref{fig:psf_sim}; due to the use of EDSER in both data sets, bin sizes shown are half of an ACIS pixel, or $0\farcs246$ on a side.

\begin{figure*}
\begin{center}
\includegraphics{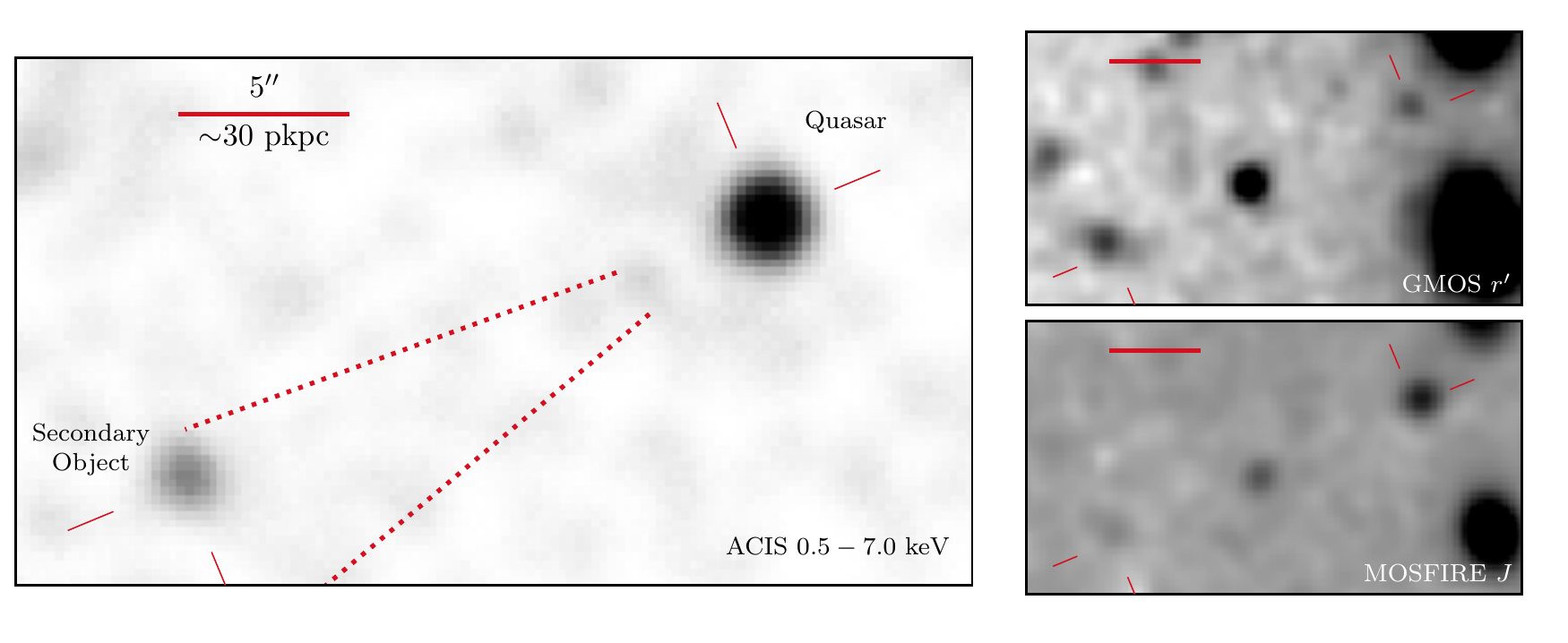}
\end{center}
\caption{Multiwavelength observations of the area around \shortname, showing the broad X-rays ({\bf left}), $r^\prime$ ({\bf top right}), and $J$ ({\bf bottom right}). All images show the same field of view, with north pointing up and east to the left, and a $5^{\prime\prime}$ scale bar is shown for reference. An object (bottom left of all panels) is seen in X-rays ${\sim}100\ {\rm kpc}$ from the quasar (top right of all panels) at the counter angle of the jets (position angles of $110^\circ$ and $130^\circ$ are indicated by the dotted lines, for reference); both objects are indicated in all images by red reticles. This source is also seen in the observed optical and NIR imaging, as discussed in the text. Additional sources present in the $r^\prime$ and $J$ images have no X-ray counterparts.
}\label{fig:multiwavelength}
\end{figure*}

To evaluate the probability of deviations from the PSF, we compare our observations to these models, quantifying deviations using the binomial probability that the flux in a bin is consistent with the expectation of the overall background plus the contribution of the PSF at that location. We follow the methods outlined by \citet{2007ApJ...657.1026W} and \citet{2014ApJ...785...17L}, namely that the probability of $N$ counts arising by chance given an expectation of $N_B$ background counts can by expressed as
\begin{equation}
P(N) = \sum_{i=N}^\infty \frac{{N_B}^i}{i!} e^{-N_B}.
\end{equation}
Here, both $N_B$ and $N$ are evaluated in a 3 by 3 bin region centered on the bin of interest.

The results of this exercise are shown in the bottom panels of Figure \ref{fig:psf_sim}. We detect no statistically significant deviation from the expectation of a quasar PSF with a stochastic background. While {\it Chandra} provides unparalleled angular resolution for X-ray observations, our technique is nevertheless insensitive to features smaller than ${\sim}4$ kpc at the redshift of \shortname\ ($0\farcs738$). In comparison, the radio structures reported by \citet{2018ApJ...861...86M} have a maximum angular extent of $0\farcs28$.

\subsection{Potential Secondary Source} \label{ssec:pot_sec}

In our initial analysis of the X-ray observations, we detected a secondary source near the position of the quasar. This source is located ${\sim}19^{\prime\prime}$ (${\sim}115$ kpc at the redshift of the quasar) from \shortname\ at a position angle of ${\sim}115^\circ$. This angle is roughly the counter-angle of the jets identified by \citet{2018ApJ...861...86M} and the extended X-ray emission reported in Section \ref{ssec:ext_str}. This source is dominated by hard energy photons, with a hardness ratio of $\mathcal{HR} = 0.0\pm0.3$ from ${\sim}20\pm5$ detected counts. {\it Chandra} observations of \shortname\ and this source are shown in the main panel of Figure \ref{fig:multiwavelength}. While there is no corresponding source in the Pan-STARRS1 imaging catalog \citep{2016arXiv161205560C}, this object is coincident with a faint source (${\sim}60 \mu{\rm Jy}$) in 3 GHz imaging \citep{2018ApJ...856L..25B}. From the deep X-ray and optical survey results of \citet{2016ApJ...817...34M}, a source with this X-ray flux and hardness and with no optical flux in Pan-STARRS1 imaging being detected  within $30^{\prime\prime}$ of \shortname\ is at least a ${\sim}2\sigma$ occurrence. 

\begin{figure*}
\begin{center}
\includegraphics{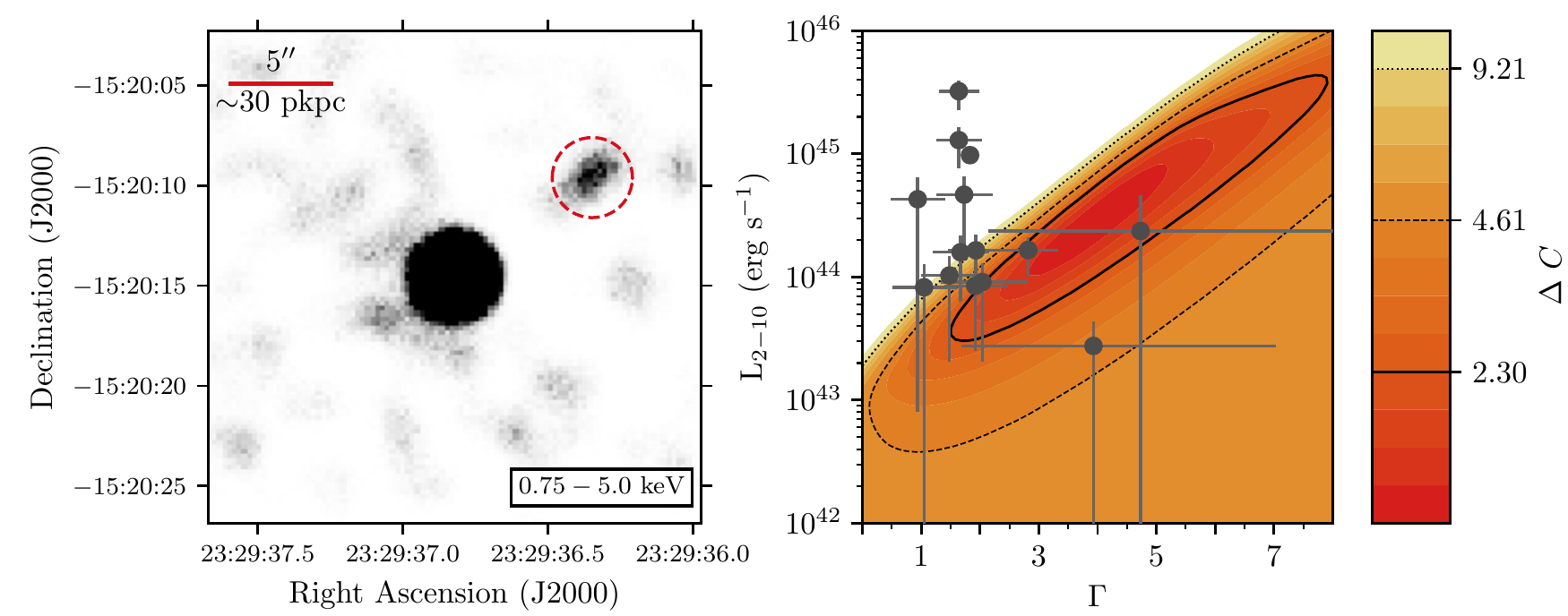}
\end{center}
\caption{{\bf Left}: Smoothed X-ray image showing the diffuse emission to the NW of \shortname\ identified in Section \ref{ssec:ext_str}. The energy range (0.75--5.0 keV), smoothing scale, and contrast have been chosen to best highlight this emission for a print image. Our $2\farcs0$ radius spectral extraction region is indicated by the dashed red circle.  
{\bf Right}: Values of $\Gamma$ and $L_{2-10}$ for the diffuse emission, colored by offset from the best-fitting $C$ value. Equivalent values to $1$, $2$, and $3\sigma$ are indicated by contour lines. A comparison sample of jets observed around intermediate redshift ($2.1 \lesssim z \lesssim 4.7$) quasars known to host radio jets from \citet{2016ApJ...833..123M} is shown with gray points.
}\label{fig:diffuse_em}
\end{figure*}

To constrain the properties of this source, we imaged the field in both the near infrared and optical regimes. For the former, we used MOSFIRE \citep{2010SPIE.7735E..1EM,2012SPIE.8446E..0JM} on the Keck-I telescope to obtain $J$-band images. We observed the field on UT 2020 July 2 for 14 52.4 second exposures (733.6 s total). Images were combined and processed with the AstrOmatic software suite \citep{1996A&AS..117..393B,2002ASPC..281..228B,2006ASPC..351..112B} following standard reductions with {\tt IRAF}. For optical imaging, we observed in the $r^\prime$ band with the GMOS-N imager \citep{2004PASP..116..425H} on the Gemini-North telescope as part of program GN-2020B-FT-101. Seven 423 second observations (2,961 s total) in a filled-hexagon dither pattern were taken in queue mode on UT 2020 July 28. Data were reduced using the DRAGONS package \citep{2019ASPC..523..321L} following standard procedures. For both sets of observations, flux calibrations were performed by comparing other objects in the field to their reported photometry in 2MASS \citep[$J$;][]{2006AJ....131.1163S} and Pan-STARRS1 \citep[$r^\prime$;][]{2016arXiv161205560C}. Photometry was extracted in $1\farcs0$ radius apertures, with backgrounds drawn in $3\farcs0 - 6\farcs0$ concentric annuli. Both observations are shown in the right of Figure \ref{fig:multiwavelength}.

At the redshift of \shortname, Ly$\alpha$ is at ${\sim}8300$ \AA, and so all emission observed in the $r^\prime$ filter will be attenuated by absorption from intervening neutral gas (the Ly$\alpha$ forest). With the small projected separation between \shortname\ and the secondary object, if they are both at $z{\approx}5.83$ we would expect the same level of attenuation to be observed for both objects. As such, the color difference between the two objects, $\Delta_{r^\prime - J}$, should only reflect the innate color differences of the two objects. If no reasonable spectral model can explain the observed values of $\Delta_{r^\prime - J}$, then the objects must be being seen through different Ly$\alpha$ forests, and we can rule out this object as being at the redshift of \shortname.

From this photometry, we find that \shortname\ has an $r^\prime - J$ color of $4.5$ mag ($r^\prime=25.8\pm0.2$, $J=21.3\pm0.1$; all magnitudes are AB). In contrast, the secondary object has a color of $r^\prime - J {=} 0.3\ {\rm mag}$ ($r^\prime=24.6\pm0.2$, $J=24.3\pm0.2$). As such, they cannot be at the same redshift and have the same spectral shape, so we rule out the possibility that this is a companion AGN. We also consider the possibility that this source is a hot spot on the counter jet, and that we are seeing rest-frame optical and UV emission from star formation. Indeed, star formation triggered on radio jets has been seen for decades \citep{1985ApJ...293...83V}, and this effect is predicted to be common in the early Universe \citep{2006ApJ...647.1040C}, although previous results have not seen star formation triggered so far from the quasar \citep{2020A&A...639L..13N}. To test this possibility, we use the \texttt{synphot} Python package to model the expected observed photometry of a quasar (using the template of \citealt{2004AJ....128..502A}) and a 
45000 K blackbody (a rough approximation of a star forming region), both at $z=5.83$. The expected color difference between a quasar and a star forming region, $\Delta_{r^\prime - J} \sim 1.5$, is not consistent with what is observed ($\Delta_{r^\prime - J} \sim 4.2$), which rules out the possibility that this object is at the redshift of the quasar. 

\section{Properties of the Extended Emission} \label{sec:ext_em_disc}

In Section \ref{ssec:ext_str} we identified extended structure in the X-ray observations to the NW of the quasar position. While without a redshift measurement we cannot conclusively confirm a connection with \shortname, it is the strongest evidence of X-ray emission from jets that we see around this quasar. Although only a limited number of photons with which to analyze this source are available, constraining the properties on this emission nevertheless sets limits on the strength of X-ray emission from one of the two radio-loudest quasars yet observed in the first billion years of the universe. To that end, we present the observed properties of the  X-ray enhancement, assuming a redshift of $z=5.831$.

We extract a spectrum from a $2\farcs0$ radius region centered on 23:29:36.35, $-$15:20:09.6. This region, which is ${\sim50}$ kpc from the centroid of the quasar emission, is shown in Figure \ref{fig:diffuse_em}. We note that, while there is a galaxy to the NW of \shortname\ visible in the optical imaging shown in Figure \ref{fig:multiwavelength}, the entirety of our extraction region is outside of the full extent of the galaxy in our $r^\prime$ and $J$ images, and no source is detected at this position in either band (to $3\sigma$ limits of $J > 24.3$, $r^\prime > 26.2$). The extraction region has $7.8^{+4.6}_{-3.5}$ net counts, with a hardness ratio of $\mathcal{HR} = 0.1^{+0.5}_{-0.4}$. We compute a binomial probability of this emission being produced by the background for three bands: soft ($p=0.020$), hard ($p=0.021$), and broad ($p=0.0015$).

\begin{figure*}
\begin{center}
\includegraphics{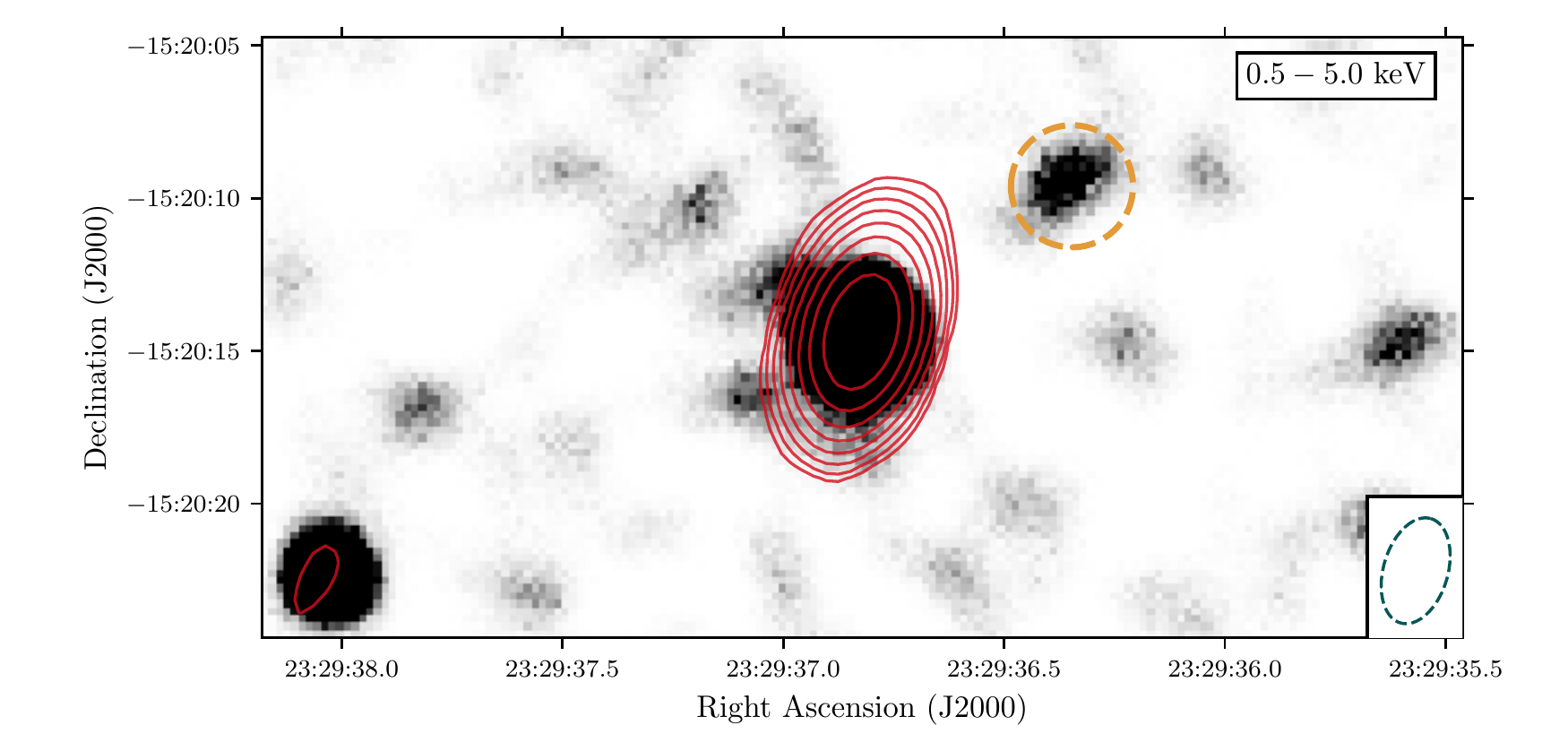}
\end{center}
\caption{Smoothed X-ray emission with contours from 3 GHz VLA observations overlaid. Contours are logarithmically spaced at levels of $f=2^n \times 9.5\ \mu{\rm Jy}\ {\rm beam}^{-1}$ for $n=1,2,3,4,5,6,7$, where $9.5\ \mu{\rm Jy}\ {\rm beam}^{-1}$ is the RMS noise of these data. The secondary object discussed in Section \ref{ssec:pot_sec} is shown in the bottom left of the figure, while the diffuse emission highlighted in Section \ref{ssec:ext_str} is indicated here by a dashed orange circle. 
Negative radio contours are not shown for clarity.
}\label{fig:radio_contours}
\end{figure*}

Using \texttt{PyXspec}, we fit the observed emission to an absorbed power law model, \texttt{phabs}$\times$\texttt{powerlaw}. The best fit of this emission comes when $\Gamma\sim4$ and $L_{2-10}\sim3\times10^{44}\ {\rm erg}\ {\rm s}^{-1}$. With $\Gamma$ and the power law normalization allowed to vary, we have two free parameters, and, with so few counts, the uncertainties on our fit are significant. Indeed, the hardness ratio, $\mathcal{HR} = 0.1^{+0.5}_{-0.4}$, is harder than that of \shortname\ ($\mathcal{HR} = -0.34^{+0.08}_{-0.09}$), implying $\Gamma < 2$. Despite the uncertainties, we do find some constraints on the observed emission. Assuming that $\Gamma{\sim}2$ \citep[e.g.,][]{2006ARA&A..44..463H}, the $1\sigma$ expectation of our fit is that this potential jet component has a luminosity of $L_{2-10}{\lesssim}10^{44}\ {\rm erg}\ {\rm s}^{-1}$, and we can conclusively rule out jet emission of $L_{2-10}{\gtrsim}10^{45}\ {\rm erg}\ {\rm s}^{-1}$. The full confidence intervals are shown in Figure \ref{fig:diffuse_em}.

As a comparison, we consider the sample of X-ray jet properties described by \citet{2016ApJ...833..123M}; this sample comes from {\it Chandra\/} imaging of eleven intermediate redshift ($2.1 \lesssim z \lesssim 4.7$) quasars known to host radio jets. In that work, \citet{2016ApJ...833..123M} fit absorbed power-laws to observed jet emission, including multiple regions for three quasars. From those fits we derive rest-frame luminosities for comparison using \texttt{XSPEC}. We note that, as the covariances between $\Gamma$ and normalization are not given for these fits, we calculated uncertainties in luminosity at the extreme values of both parameters, and therefore these uncertainties are most likely overestimates. The observed properties of these jets are shown in the right panel of Figure \ref{fig:diffuse_em}.

Assuming that this emission is not driven by a source with an extreme power law spectral index ($\Gamma \gtrsim 5$), \shortname\ is not host to the most X-ray luminous jets in the $z>2$ universe. In contrast, for a steeper spectral slope, the high predicted luminosity is caused by extrapolating a poorly-constrained spectrum into unobserved energy ranges \citep[e.g.,][]{2020ApJ...900..189C} and is not indicative of an expected relation between $\Gamma$ and $L_{2-10}$. As can be seen in Figure \ref{fig:delta_aox}, \shortname\ is not an outlier in radio loudness among the broad population of quasars; the only reason we would expect this emission to be large is due to the $(1+z)^4$ scaling of the CMB energy density.

Following \citet{2017MNRAS.466.4299L}, we calculate an equivalent magnetic field to the CMB's energy density as
\begin{equation}
B_{\rm CMB} = 3.26 (1+z)^{2} \Gamma_{\rm jet}\, \mu{\rm G},
\end{equation}
where $\Gamma_{\rm jet}$ is the bulk Lorentz factor, finding $B_{\rm CMB} = 152\, \Gamma_{\rm jet}\, \mu {\rm G}$ at this redshift. If magnetic field strengths inferred from radio synchrotron measurements are less than this value then particle cooling should be dominated by X-ray inverse Compton processes. While values of $\Gamma_{\rm jet}\sim10$ have been seen at $z>5$ \citep[e.g.,][]{2015MNRAS.446.2921F}, the VLBI analysis of \citet{2018ApJ...861...86M} did not find evidence for such fast outflows in this system, and so we adopt a conservative $\Gamma_{\rm jet}\sim1$. For a jet of relativistic particles with X-ray emission caused by interaction with the CMB at redshift $z$ and radio emission driven by a magnetic field of strength $B$, the relative X-ray and radio flux densities can be found through
\begin{equation}\label{eqn:radio_xray_flux_to_B}
\left( \frac{S_{r}\nu_r}{S_X \nu_X} \right)= \frac{1}{2.89} \left(\frac{B}{\mu {\rm G}} \right)^2 (1 + z)^{-4},
\end{equation}
where $S_X$ and $S_r$ are the X-ray and radio flux densities at observed frequencies $\nu_X$ and $\nu_r$, respectively, and $S_r \propto \nu$ is assumed \citep{2002ARA&A..40..319C}.

To constrain the radio emission, we examine archival observations of \shortname\ taken with NSF's Karl G. Jansky Very Large Array (VLA) at 3 GHz in January 2018 \citep{2018ApJ...861L..14B}. We reprocessed these observations using standard techniques \citep[e.g.,][]{2019ApJ...874L..32C}. The resulting image, shown in Figure \ref{fig:radio_contours}, has an RMS noise of $9.5\ \mu{\rm Jy}\ {\rm beam}^{-1}$ and a resolution of $2\farcs6 \times 1\farcs4$ (${\rm PA} = -13^\circ$). The secondary source to the SE of the quasar discussed in Section \ref{ssec:pot_sec} has a nominal radio counterpart of $60 \mu{\rm Jy}\ {\rm beam}^{-1}$ (${\sim}6\sigma$ significance). At the location of the diffuse X-ray emission being discussed here, no radio emission is seen.
If we assume this X-ray emission is from inverse Compton up-scattering of the CMB by relativistic electrons, and adopting flux densities of $<28.5 \mu{\rm Jy}$ ($3\sigma$) at 3 GHz and $58^{+55}_{-43} \times 10^{-12}$ Jy at $2.4\times 10^{17}$ Hz (1 keV; from the normalization of the X-ray spectral fit), then from Equation \ref{eqn:radio_xray_flux_to_B} we find an expected source magnetic field of ${\lesssim}6\ \mu\textrm{G}$. This is a typical magnetic field strength for diffuse lobes of extragalactic radio sources \citep{1980ARA&A..18..165M} and is significantly lower than the value of ${\sim}3.5$ mG reported for the inner kpc of the quasar by \citet{2018ApJ...861...86M}. Deeper radio observations are required to test if there is a radio counterpart to this possible X-ray knot around \shortname.

\section{Discussion} \label{sec:discussion}

In this work, we have presented analysis of X-ray observations of \shortname. In addition to the X-ray properties of the quasar itself, we also conducted a search for evidence of IC/CMB emission. Here, we discuss our results in the context of the broader perspective of high-redshift SMBH growth, AGN emission, and radio activity around quasars

One of the primary goals of this investigation was to search for evidence of extended jets as an indicator of enhanced accretion rates for high redshift quasars. \citet{2008MNRAS.386..989J} presented a model (see also \citealt{2013MNRAS.432.2818G} and \citealt{2009MNRAS.400.1521J}) wherein a magnetic torque on the accretion disk extracts energy that is then injected into a magnetized jet. As the jet transports a minimal amount of mass but a more substantial amount of angular momentum, the net effect of this model is that the jet enhances the mass accretion rate, allowing for more significant growth than would be possible without the jet. 
Here, the model postulates that the jet is powered by extraction of energy and momentum from the disk (the Blandford--Payne processes; \citealt{1982MNRAS.199..883B}) rather than from the spin-energy of the black hole (the Blandford--Znajek mechanism; \citealt{1977MNRAS.179..433B}), as is commonly invoked at lower redshifts. Because of this distinction, observed relations showing a decline in accretion rate with radio loudness \citep[e.g.,][]{2007ApJ...658..815S} driven by conditions where jets are powered from the black hole spin \citep[e.g.,][]{2013ApJ...765...62S} are not pertinent to the ability of jets to enhance accretion rates in this model.

Potentially, this model of jet-enhanced accretion could complement the jet emission models of \citet{2010MNRAS.406..975G}, who present an evolutionary model for radio-loud AGN, starting with retrograde spins that produce Blandford--Payne jets and transitioning to Blandford--Znajek-dominated prograde systems. Prograde systems are more efficient at converting accreting mass to jet energy \citep{2012MNRAS.423L..55T}, so the presence of large jets around a retrograde SMBH could indicate significant amounts of accretion have occurred. As black hole spin measurements at these redshifts are beyond the capabilities of current observatories (although see \citealt{2015AN....336..312G}, who postulate that ULAS J1120+0641 at $z=7.08$ has a retrograde spin), a direct test of this hypothesis is not currently possible. Nevertheless, this scenario does present a case in which jet emission could enable the rapid buildup of SMBHs needed to reconcile observed quasar populations with theoretical models of seed formation.

Simulations provide some context to the role of jet feedback in regulating AGN growth in the early universe. Recently, \citet{2019MNRAS.486.3892R} simulated the formation of an SMBH through a metal-free supermassive star that directly collapses into a ${\sim}10^4\ M_{\odot}$ black hole. Immediately after the collapse, mechanical feedback is turned on in the form of jets; these jets suppress accretion, driving the black hole growth to sub-Eddington values, but their effect is limited to the inner 0.1 pc. In contrast, \citet{2020MNRAS.497..302T} simulated both radiative and mechanical feedback from a black hole of initial mass $10^5\ M_{\odot}$ and found that, after around one dynamical timescale (${\sim} 8.4 \times 10^{6}$ yr), a cascade of neutral gas along the equatorial region drives major outflows in the polar region. The balance found in this simulation is one of hyper-Eddington accretion and an evacuation of gas along the poles.

In contrast, the VLBI observations of jets presented by \citet{2018ApJ...861...86M} provide direct evidence that the jets around \shortname\ have, at least, extended beyond a kpc from the SMBH itself. From the candidate jet emission reported here, the jets could have extended to ${\sim 50}$ kpc, and the implied duration of jet launching is, assuming $v\sim0.3 c$, of order 1 Myr. Recent interferometric observations of the $z=6.1$ blazar PSO J0309+27 by \citet{2020A&A...643L..12S} showed jets with projected sizes of hundreds of parsecs; while these structures are potentially larger given the viewing angle, they, in addition to the observations of \citet{2018ApJ...861...86M}, are in contrast to the limited effect of jets predicted by \citet{2019MNRAS.486.3892R} and further support the notion that these jets could be aiding in accretion.

Recently, \citet{2020ApJ...904...57S} reported two candidate X-ray jets at intermediate redshifts ($z\sim3.2$) found using a technique similar to that employed here in Section \ref{ssec:ext_str}. Although these jets are along a line connecting quasar cores to radio emission, they have no associated radio emission of their own. Similarly, X-ray emission associated with a $z\sim4.3$ quasar reported by \citet{2003ApJ...598L..15S} has no corresponding radio emission. For the three other jets, the X-ray luminosities are of order ${\rm L}_{2-10}\sim10^{45}\ {\rm erg}\ {\rm s}^{-1}$, slightly higher than what we found here. While individually less compelling, the combination of all four reports of X-ray emission paints a picture wherein jets around intermediate and high-redshift quasars may be hiding without associated detections of radio emission. If this is the case, only deep X-ray observations of radio-loud high-redshift quasars will enable a meaningful insight into this population.

\section{Summary} \label{sec:summary}

We have presented deep (265 ks) X-ray observations of \shortname, one of the two radio loudest quasars observed in the first billion years of the universe \citep[$z>5.6$; see also][]{2020A&A...635L...7B}. As part of our analysis, we have also obtained optical and NIR imaging of the quasar and its surroundings. Our primary results are
\begin{itemize}
    \item We fit the X-ray properties of the quasar itself, which is detected with over 100 counts; adopting an absorbed power law model, we find the AGN emission is fit with $\Gamma = 1.99^{+0.29}_{-0.28}$, and it has a corresponding unobscured luminosity of $L_{2-10} = 1.26^{+0.45}_{-0.33} \times 10^{45}\ {\rm erg}\ {\rm s}^{-1}$. 
    \item In relation to the quasar's rest-frame UV emission, the observed X-ray luminosity is stronger than expected from the scaling relation for radio-quiet quasars of \citet{2016ApJ...819..154L}, with $\Delta \alpha_{\rm OX} = 0.25 \pm 0.11$. However, in comparison to other radio-loud quasars, \shortname\ is not an outlier in its X-ray loudness.
    \item As radio jets have been observed around \shortname\ \citep{2018ApJ...861...86M}, we searched for evidence of these jets in X-rays. We found no excess X-ray emission in the core of the quasar, either in excess luminosity or in deviations from the expected PSF. We also investigated a nearby (${\sim}19^{\prime\prime}$ offset) X-ray source along the predicted jet axis; followup optical and NIR imaging presented here rule out this emission being at the redshift of \shortname.
    \item Through an analysis of the angular overdensity of emission, we identified a peak in the X-ray emission along the observed jet axis, corresponding to a $\gtrsim3\sigma$ deviation above the expectation. Further analysis of a $2\farcs0$ radius region finds that this emission is unlikely to be associated with a background fluctuation, with a binomial probability of $P=0.0015$. As this emission is at the same position angle as the previously reported radio emission, we take this as tentative evidence for X-ray emission from the jets ${\sim}50$ kpc from the position of the quasar.
    \item Spectral analysis of the jet emission is limited by the small number of observed photons in the extraction region. However, assuming a spectral index similar to that found for other jet emission, the overall luminosity is $L_{2-10}\lesssim10^{44}\ {\rm erg}\ {\rm s}^{-1}$. 
    \item The tentative detection of X-ray emission along the jet axis of \shortname\ with no radio or optical counterpart is potentially indicative that this emission is being produced by inverse Compton interactions of jetted particles with the Cosmic Microwave Background. Further, deeper observations are required to fully test this possibility.
\end{itemize}

The depth of the observations required to obtain these results -- and the uncertainties still remaining -- provide further support for the need for a new generation of X-ray observatories, namely {\it Athena} \citep{2013arXiv1306.2307N} and, hopefully, {\it Lynx} \citep{2019JATIS...5b1001G} and {\it AXIS} \citep{2019BAAS...51g.107M}.

%\acknowledgments
\vspace{2mm}

{\small The work of T.C. and D.S. was carried out at the Jet Propulsion Laboratory, California Institute of Technology, under a contract with NASA. T.C.'s research was supported by an appointment to the NASA Postdoctoral Program at the Jet Propulsion Laboratory, California Institute of Technology, administered by Universities Space Research Association under contract with NASA. 

The scientific results reported in this article are based on observations made by the \textit{Chandra X-ray Observatory}. This research has made use of software provided by the \textit{Chandra} X-ray Center (CXC) in the application package CIAO. Based on observations obtained at the international Gemini Observatory (GN-2020B-FT-101), a program of NSF’s NOIRLab, which is managed by the Association of Universities for Research in Astronomy (AURA) under a cooperative agreement with the National Science Foundation on behalf of the Gemini Observatory partnership: the National Science Foundation (United States), National Research Council (Canada), Agencia Nacional de Investigaci\'{o}n y Desarrollo (Chile), Ministerio de Ciencia, Tecnolog\'{i}a e Innovaci\'{o}n (Argentina), Minist\'{e}rio da Ci\^{e}ncia, Tecnologia, Inova\c{c}\~{o}es e Comunica\c{c}\~{o}es (Brazil), and Korea Astronomy and Space Science Institute (Republic of Korea). Some of the data presented herein were obtained at the W. M. Keck Observatory, which is operated as a scientific partnership among the California Institute of Technology, the University of California and the National Aeronautics and Space Administration. The Observatory was made possible by the generous financial support of the W. M. Keck Foundation. 

This work was enabled by observations made from the Keck and Gemini North telescopes, located within the Maunakea Science Reserve and adjacent to the summit of Maunakea. The authors wish to recognize and acknowledge the very significant cultural role and reverence that the summit of Maunakea has always had within the indigenous Hawaiian community.  We are most fortunate to have the opportunity to conduct observations from this mountain and are grateful for the privilege of observing the Universe from a place that is unique in both its astronomical quality and its cultural significance.

}

\vspace{5mm}
\facility{CXO,
        Gemini:Gillett (GMOS-N),
        Keck:I (MOSFIRE)}
\software{BEHR \citep{2006ApJ...652..610P},
          CIAO \citep{2006SPIE.6270E..1VF},
          MARX \citep{2012SPIE.8443E..1AD},
          PyFITS \citep{1999ASPC..172..483B},
          Synphot \citep{2018ascl.soft11001S},
          WAVDETECT \citep{2002ApJS..138..185F},
          XSPEC \citep{1996ASPC..101...17A}}

\textcopyright\ 2020. All rights reserved.


\begin{thebibliography}{}
\providecommand\natexlab[1]{#1}
\providecommand\JournalTitle[1]{#1}

\bibitem[{{Abazajian} {et~al.}(2004){Abazajian}, {Adelman-McCarthy},
  {Ag{\"u}eros}, {Allam}, {Anderson}, {Anderson}, {Annis}, {Bahcall}, {Baldry},
  {Bastian}, {Berlind}, {Bernardi}, {Blanton}, {Bochanski}, {Boroski},
  {Briggs}, {Brinkmann}, {Brunner}, {Budav{\'a}ri}, {Carey}, {Carliles},
  {Castander}, {Connolly}, {Csabai}, {Doi}, {Dong}, {Eisenstein}, {Evans},
  {Fan}, {Finkbeiner}, {Friedman}, {Frieman}, {Fukugita}, {Gal}, {Gillespie},
  {Glazebrook}, {Gray}, {Grebel}, {Gunn}, {Gurbani}, {Hall}, {Hamabe},
  {Harris}, {Harris}, {Harvanek}, {Heckman}, {Hendry}, {Hennessy}, {Hindsley},
  {Hogan}, {Hogg}, {Holmgren}, {Ichikawa}, {Ichikawa}, {Ivezi{\'c}}, {Jester},
  {Johnston}, {Jorgensen}, {Kent}, {Kleinman}, {Knapp}, {Kniazev}, {Kron},
  {Krzesinski}, {Kunszt}, {Kuropatkin}, {Lamb}, {Lampeitl}, {Lee}, {Leger},
  {Li}, {Lin}, {Loh}, {Long}, {Loveday}, {Lupton}, {Malik}, {Margon},
  {Matsubara}, {McGehee}, {McKay}, {Meiksin}, {Munn}, {Nakajima}, {Nash},
  {Neilsen}, {Newberg}, {Newman}, {Nichol}, {Nicinski}, {Nieto-Santisteban},
  {Nitta}, {Okamura}, {O'Mullane}, {Ostriker}, {Owen}, {Padmanabhan},
  {Peoples}, {Pier}, {Pope}, {Quinn}, {Richards}, {Richmond}, {Rix}, {Rockosi},
  {Schlegel}, {Schneider}, {Scranton}, {Sekiguchi}, {Seljak}, {Sergey},
  {Sesar}, {Sheldon}, {Shimasaku}, {Siegmund}, {Silvestri}, {Smith},
  {Smol{\v{c}}i{\'c}}, {Snedden}, {Stebbins}, {Stoughton}, {Strauss},
  {SubbaRao}, {Szalay}, {Szapudi}, {Szkody}, {Szokoly}, {Tegmark}, {Teodoro},
  {Thakar}, {Tremonti}, {Tucker}, {Uomoto}, {Vanden Berk}, {Vandenberg},
  {Vogeley}, {Voges}, {Vogt}, {Walkowicz}, {Wang}, {Weinberg}, {West}, {White},
  {Wilhite}, {Xu}, {Yanny}, {Yasuda}, {Yip}, {Yocum}, {York}, {Zehavi},
  {Zibetti}, \& {Zucker}}]{2004AJ....128..502A}
{Abazajian}, K., {Adelman-McCarthy}, J.~K., {Ag{\"u}eros}, M.~A., {et~al.}
  2004, \href{http://dx.doi.org/10.1086/421365}{\JournalTitle{\aj}, 128, 502}

\bibitem[{{Arnaud}(1996)}]{1996ASPC..101...17A}
{Arnaud}, K.~A. 1996, \JournalTitle{{adass V}}, 101, 17

\bibitem[{{Ba{\~n}ados} {et~al.}(2018{\natexlab{a}}){Ba{\~n}ados}, {Carilli},
  {Walter}, {Momjian}, {Decarli}, {Farina}, {Mazzucchelli}, \&
  {Venemans}}]{2018ApJ...861L..14B}
{Ba{\~n}ados}, E., {Carilli}, C., {Walter}, F., {et~al.} 2018{\natexlab{a}},
  \href{http://dx.doi.org/10.3847/2041-8213/aac511}{\JournalTitle{\apjl}, 861,
  L14}

\bibitem[{{Ba{\~n}ados} {et~al.}(2015){Ba{\~n}ados}, {Venemans}, {Morganson},
  {Hodge}, {Decarli}, {Walter}, {Stern}, {Schlafly}, {Farina}, {Greiner},
  {Chambers}, {Fan}, {Rix}, {Burgett}, {Draper}, {Flewelling}, {Kaiser},
  {Metcalfe}, {Morgan}, {Tonry}, \& {Wainscoat}}]{2015ApJ...804..118B}
{Ba{\~n}ados}, E., {Venemans}, B.~P., {Morganson}, E., {et~al.} 2015,
  \href{http://dx.doi.org/10.1088/0004-637X/804/2/118}{\JournalTitle{\apj},
  804, 118}

\bibitem[{{Ba{\~n}ados} {et~al.}(2016){Ba{\~n}ados}, {Venemans}, {Decarli},
  {Farina}, {Mazzucchelli}, {Walter}, {Fan}, {Stern}, {Schlafly}, {Chambers},
  {Rix}, {Jiang}, {McGreer}, {Simcoe}, {Wang}, {Yang}, {Morganson}, {De Rosa},
  {Greiner}, {Balokovi{\'c}}, {Burgett}, {Cooper}, {Draper}, {Flewelling},
  {Hodapp}, {Jun}, {Kaiser}, {Kudritzki}, {Magnier}, {Metcalfe}, {Miller},
  {Schindler}, {Tonry}, {Wainscoat}, {Waters}, \& {Yang}}]{2016ApJS..227...11B}
{Ba{\~n}ados}, E., {Venemans}, B.~P., {Decarli}, R., {et~al.} 2016,
  \href{http://dx.doi.org/10.3847/0067-0049/227/1/11}{\JournalTitle{\apjs},
  227, 11}

\bibitem[{{Ba{\~n}ados} {et~al.}(2018{\natexlab{b}}){Ba{\~n}ados}, {Venemans},
  {Mazzucchelli}, {Farina}, {Walter}, {Wang}, {Decarli}, {Stern}, {Fan},
  {Davies}, {Hennawi}, {Simcoe}, {Turner}, {Rix}, {Yang}, {Kelson}, {Rudie}, \&
  {Winters}}]{2018Natur.553..473B}
{Ba{\~n}ados}, E., {Venemans}, B.~P., {Mazzucchelli}, C., {et~al.}
  2018{\natexlab{b}},
  \href{http://dx.doi.org/10.1038/nature25180}{\JournalTitle{\nat}, 553, 473}

\bibitem[{{Ba{\~n}ados} {et~al.}(2018{\natexlab{c}}){Ba{\~n}ados}, {Connor},
  {Stern}, {Mulchaey}, {Fan}, {Decarli}, {Farina}, {Mazzucchelli}, {Venemans},
  {Walter}, {Wang}, \& {Yang}}]{2018ApJ...856L..25B}
{Ba{\~n}ados}, E., {Connor}, T., {Stern}, D., {et~al.} 2018{\natexlab{c}},
  \href{http://dx.doi.org/10.3847/2041-8213/aab61e}{\JournalTitle{\apjl}, 856,
  L25}

\bibitem[{{Barrett} \& {Bridgman}(1999)}]{1999ASPC..172..483B}
{Barrett}, P.~E., \& {Bridgman}, W.~T. 1999, \JournalTitle{{adass VIII}}, 172,
  483

\bibitem[{{Belladitta} {et~al.}(2020){Belladitta}, {Moretti}, {Caccianiga},
  {Spingola}, {Severgnini}, {Della Ceca}, {Ghisellini}, {Dallacasa},
  {Sbarrato}, {Cicone}, {Cassar{\`a}}, \& {Pedani}}]{2020A&A...635L...7B}
{Belladitta}, S., {Moretti}, A., {Caccianiga}, A., {et~al.} 2020,
  \href{http://dx.doi.org/10.1051/0004-6361/201937395}{\JournalTitle{\aap},
  635, L7}

\bibitem[{{Bertin}(2006)}]{2006ASPC..351..112B}
{Bertin}, E. 2006, \JournalTitle{adass XV}, 351, 112

\bibitem[{{Bertin} \& {Arnouts}(1996)}]{1996A&AS..117..393B}
{Bertin}, E., \& {Arnouts}, S. 1996,
  \href{http://dx.doi.org/10.1051/aas:1996164}{\JournalTitle{\aaps}, 117, 393}

\bibitem[{{Bertin} {et~al.}(2002){Bertin}, {Mellier}, {Radovich}, {Missonnier},
  {Didelon}, \& {Morin}}]{2002ASPC..281..228B}
{Bertin}, E., {Mellier}, Y., {Radovich}, M., {et~al.} 2002, \JournalTitle{adass
  XI}, 281, 228

\bibitem[{{Blandford} \& {Payne}(1982)}]{1982MNRAS.199..883B}
{Blandford}, R.~D., \& {Payne}, D.~G. 1982,
  \href{http://dx.doi.org/10.1093/mnras/199.4.883}{\JournalTitle{\mnras}, 199,
  883}

\bibitem[{{Blandford} \& {Znajek}(1977)}]{1977MNRAS.179..433B}
{Blandford}, R.~D., \& {Znajek}, R.~L. 1977,
  \href{http://dx.doi.org/10.1093/mnras/179.3.433}{\JournalTitle{\mnras}, 179,
  433}

\bibitem[{{Breiding} {et~al.}(2017){Breiding}, {Meyer}, {Georganopoulos},
  {Keenan}, {DeNigris}, \& {Hewitt}}]{2017ApJ...849...95B}
{Breiding}, P., {Meyer}, E.~T., {Georganopoulos}, M., {et~al.} 2017,
  \href{http://dx.doi.org/10.3847/1538-4357/aa907a}{\JournalTitle{\apj}, 849,
  95}

\bibitem[{{Brightman} {et~al.}(2013){Brightman}, {Silverman}, {Mainieri},
  {Ueda}, {Schramm}, {Matsuoka}, {Nagao}, {Steinhardt}, {Kartaltepe},
  {Sanders}, {Treister}, {Shemmer}, {Brandt}, {Brusa}, {Comastri}, {Ho},
  {Lanzuisi}, {Lusso}, {Nandra}, {Salvato}, {Zamorani}, {Akiyama}, {Alexander},
  {Bongiorno}, {Capak}, {Civano}, {Del Moro}, {Doi}, {Elvis}, {Hasinger},
  {Laird}, {Masters}, {Mignoli}, {Ohta}, {Schawinski}, \&
  {Taniguchi}}]{2013MNRAS.433.2485B}
{Brightman}, M., {Silverman}, J.~D., {Mainieri}, V., {et~al.} 2013,
  \href{http://dx.doi.org/10.1093/mnras/stt920}{\JournalTitle{\mnras}, 433,
  2485}

\bibitem[{{Carilli} {et~al.}(2019){Carilli}, {Perley}, {Dhawan}, \&
  {Perley}}]{2019ApJ...874L..32C}
{Carilli}, C.~L., {Perley}, R.~A., {Dhawan}, V., \& {Perley}, D.~A. 2019,
  \href{http://dx.doi.org/10.3847/2041-8213/ab1019}{\JournalTitle{\apjl}, 874,
  L32}

\bibitem[{{Carilli} \& {Taylor}(2002)}]{2002ARA&A..40..319C}
{Carilli}, C.~L., \& {Taylor}, G.~B. 2002,
  \href{http://dx.doi.org/10.1146/annurev.astro.40.060401.093852}{\JournalTitle{\araa},
  40, 319}

\bibitem[{{Cash}(1979)}]{1979ApJ...228..939C}
{Cash}, W. 1979, \href{http://dx.doi.org/10.1086/156922}{\JournalTitle{\apj},
  228, 939}

\bibitem[{{Chambers} {et~al.}(2016){Chambers}, {Magnier}, {Metcalfe},
  {Flewelling}, {Huber}, {Waters}, {Denneau}, {Draper}, {Farrow}, {Finkbeiner},
  {Holmberg}, {Koppenhoefer}, {Price}, {Rest}, {Saglia}, {Schlafly}, {Smartt},
  {Sweeney}, {Wainscoat}, {Burgett}, {Chastel}, {Grav}, {Heasley}, {Hodapp},
  {Jedicke}, {Kaiser}, {Kudritzki}, {Luppino}, {Lupton}, {Monet}, {Morgan},
  {Onaka}, {Shiao}, {Stubbs}, {Tonry}, {White}, {Ba{\~n}ados}, {Bell},
  {Bender}, {Bernard}, {Boegner}, {Boffi}, {Botticella}, {Calamida},
  {Casertano}, {Chen}, {Chen}, {Cole}, {Deacon}, {Frenk}, {Fitzsimmons},
  {Gezari}, {Gibbs}, {Goessl}, {Goggia}, {Gourgue}, {Goldman}, {Grant},
  {Grebel}, {Hambly}, {Hasinger}, {Heavens}, {Heckman}, {Henderson}, {Henning},
  {Holman}, {Hopp}, {Ip}, {Isani}, {Jackson}, {Keyes}, {Koekemoer}, {Kotak},
  {Le}, {Liska}, {Long}, {Lucey}, {Liu}, {Martin}, {Masci}, {McLean}, {Mindel},
  {Misra}, {Morganson}, {Murphy}, {Obaika}, {Narayan}, {Nieto-Santisteban},
  {Norberg}, {Peacock}, {Pier}, {Postman}, {Primak}, {Rae}, {Rai}, {Riess},
  {Riffeser}, {Rix}, {R{\"o}ser}, {Russel}, {Rutz}, {Schilbach}, {Schultz},
  {Scolnic}, {Strolger}, {Szalay}, {Seitz}, {Small}, {Smith}, {Soderblom},
  {Taylor}, {Thomson}, {Taylor}, {Thakar}, {Thiel}, {Thilker}, {Unger},
  {Urata}, {Valenti}, {Wagner}, {Walder}, {Walter}, {Watters}, {Werner},
  {Wood-Vasey}, \& {Wyse}}]{2016arXiv161205560C}
{Chambers}, K.~C., {Magnier}, E.~A., {Metcalfe}, N., {et~al.} 2016,
  \JournalTitle{arXiv e-prints}, arXiv:1612.05560

\bibitem[{{Chartas} {et~al.}(2000){Chartas}, {Worrall}, {Birkinshaw},
  {Cresitello-Dittmar}, {Cui}, {Ghosh}, {Harris}, {Hooper}, {Jauncey}, {Kim},
  {Lovell}, {Mathur}, {Schwartz}, {Tingay}, {Virani}, \&
  {Wilkes}}]{2000ApJ...542..655C}
{Chartas}, G., {Worrall}, D.~M., {Birkinshaw}, M., {et~al.} 2000,
  \href{http://dx.doi.org/10.1086/317049}{\JournalTitle{\apj}, 542, 655}

\bibitem[{{Connor} {et~al.}(2018){Connor}, {Kelson}, {Mulchaey}, {Vikhlinin},
  {Patel}, {Balogh}, {Joshi}, {Kraft}, {Nagai}, \&
  {Starikova}}]{2018ApJ...867...25C}
{Connor}, T., {Kelson}, D.~D., {Mulchaey}, J., {et~al.} 2018,
  \href{http://dx.doi.org/10.3847/1538-4357/aae38b}{\JournalTitle{\apj}, 867,
  25}

\bibitem[{{Connor} {et~al.}(2019){Connor}, {Ba{\~n}ados}, {Stern}, {Decarli},
  {Schindler}, {Fan}, {Farina}, {Mazzucchelli}, {Mulchaey}, \&
  {Walter}}]{2019ApJ...887..171C}
{Connor}, T., {Ba{\~n}ados}, E., {Stern}, D., {et~al.} 2019,
  \href{http://dx.doi.org/10.3847/1538-4357/ab5585}{\JournalTitle{\apj}, 887,
  171}

\bibitem[{{Connor} {et~al.}(2020){Connor}, {Ba{\~n}ados}, {Mazzucchelli},
  {Stern}, {Decarli}, {Fan}, {Farina}, {Lusso}, {Neeleman}, \&
  {Walter}}]{2020ApJ...900..189C}
{Connor}, T., {Ba{\~n}ados}, E., {Mazzucchelli}, C., {et~al.} 2020,
  \href{http://dx.doi.org/10.3847/1538-4357/abaab9}{\JournalTitle{\apj}, 900,
  189}

\bibitem[{{Croft} {et~al.}(2006){Croft}, {van Breugel}, {de Vries}, {Dopita},
  {Martin}, {Morganti}, {Neff}, {Oosterloo}, {Schiminovich}, {Stanford}, \&
  {van Gorkom}}]{2006ApJ...647.1040C}
{Croft}, S., {van Breugel}, W., {de Vries}, W., {et~al.} 2006,
  \href{http://dx.doi.org/10.1086/505526}{\JournalTitle{\apj}, 647, 1040}

\bibitem[{{Davis} {et~al.}(2012){Davis}, {Bautz}, {Dewey}, {Heilmann}, {Houck},
  {Huenemoerder}, {Marshall}, {Nowak}, {Schattenburg}, {Schulz}, \&
  {Smith}}]{2012SPIE.8443E..1AD}
{Davis}, J.~E., {Bautz}, M.~W., {Dewey}, D., {et~al.} 2012,
  \href{http://dx.doi.org/10.1117/12.926937}{\JournalTitle{Proc. SPIE}, 8443,
  84431A}

\bibitem[{{Fabian}(2016)}]{2016AN....337..375F}
{Fabian}, A.~C. 2016,
  \href{http://dx.doi.org/10.1002/asna.201612316}{\JournalTitle{AN}, 337, 375}

\bibitem[{{Fabian} {et~al.}(2014){Fabian}, {Walker}, {Celotti}, {Ghisellini},
  {Mocz}, {Blundell}, \& {McMahon}}]{2014MNRAS.442L..81F}
{Fabian}, A.~C., {Walker}, S.~A., {Celotti}, A., {et~al.} 2014,
  \href{http://dx.doi.org/10.1093/mnrasl/slu065}{\JournalTitle{\mnras}, 442,
  L81}

\bibitem[{{Freeman} {et~al.}(2002){Freeman}, {Kashyap}, {Rosner}, \&
  {Lamb}}]{2002ApJS..138..185F}
{Freeman}, P.~E., {Kashyap}, V., {Rosner}, R., \& {Lamb}, D.~Q. 2002,
  \href{http://dx.doi.org/10.1086/324017}{\JournalTitle{\apjs}, 138, 185}

\bibitem[{{Frey} {et~al.}(2015){Frey}, {Paragi}, {Fogasy}, \&
  {Gurvits}}]{2015MNRAS.446.2921F}
{Frey}, S., {Paragi}, Z., {Fogasy}, J.~O., \& {Gurvits}, L.~I. 2015,
  \href{http://dx.doi.org/10.1093/mnras/stu2294}{\JournalTitle{\mnras}, 446,
  2921}

\bibitem[{{Fruscione} {et~al.}(2006){Fruscione}, {McDowell}, {Allen},
  {Brickhouse}, {Burke}, {Davis}, {Durham}, {Elvis}, {Galle}, {Harris},
  {Huenemoerder}, {Houck}, {Ishibashi}, {Karovska}, {Nicastro}, {Noble},
  {Nowak}, {Primini}, {Siemiginowska}, {Smith}, \&
  {Wise}}]{2006SPIE.6270E..1VF}
{Fruscione}, A., {McDowell}, J.~C., {Allen}, G.~E., {et~al.} 2006,
  \href{http://dx.doi.org/10.1117/12.671760}{\JournalTitle{Proc. SPIE}, 6270,
  62701V}

\bibitem[{{Garmire} {et~al.}(2003){Garmire}, {Bautz}, {Ford}, {Nousek}, \&
  {Ricker}}]{2003SPIE.4851...28G}
{Garmire}, G.~P., {Bautz}, M.~W., {Ford}, P.~G., {Nousek}, J.~A., \& {Ricker},
  George~R., J. 2003,
  \href{http://dx.doi.org/10.1117/12.461599}{\JournalTitle{Proc. SPIE}, 4851,
  28}

\bibitem[{{Garofalo} {et~al.}(2010){Garofalo}, {Evans}, \&
  {Sambruna}}]{2010MNRAS.406..975G}
{Garofalo}, D., {Evans}, D.~A., \& {Sambruna}, R.~M. 2010,
  \href{http://dx.doi.org/10.1111/j.1365-2966.2010.16797.x}{\JournalTitle{\mnras},
  406, 975}

\bibitem[{{Gaskin} {et~al.}(2019){Gaskin}, {Swartz}, {Vikhlinin}, {{\"O}zel},
  {Gelmis}, {Arenberg}, {Bandler}, {Bautz}, {Civitani}, {Dominguez}, {Eckart},
  {Falcone}, {Figueroa-Feliciano}, {Freeman}, {G{\"u}nther}, {Havey},
  {Heilmann}, {Kilaru}, {Kraft}, {McCarley}, {McEntaffer}, {Pareschi},
  {Purcell}, {Reid}, {Schattenburg}, {Schwartz}, {Schwartz}, {Tananbaum},
  {Tremblay}, {Zhang}, \& {Zuhone}}]{2019JATIS...5b1001G}
{Gaskin}, J.~A., {Swartz}, D.~A., {Vikhlinin}, A., {et~al.} 2019,
  \href{http://dx.doi.org/10.1117/1.JATIS.5.2.021001}{\JournalTitle{JATIS}, 5,
  021001}

\bibitem[{{Gehrels}(1986)}]{1986ApJ...303..336G}
{Gehrels}, N. 1986,
  \href{http://dx.doi.org/10.1086/164079}{\JournalTitle{\apj}, 303, 336}

\bibitem[{{Ghisellini} {et~al.}(2015){Ghisellini}, {Haardt}, {Ciardi},
  {Sbarrato}, {Gallo}, {Tavecchio}, \& {Celotti}}]{2015MNRAS.452.3457G}
{Ghisellini}, G., {Haardt}, F., {Ciardi}, B., {et~al.} 2015,
  \href{http://dx.doi.org/10.1093/mnras/stv1541}{\JournalTitle{\mnras}, 452,
  3457}

\bibitem[{{Ghisellini} {et~al.}(2013){Ghisellini}, {Haardt}, {Della Ceca},
  {Volonteri}, \& {Sbarrato}}]{2013MNRAS.432.2818G}
{Ghisellini}, G., {Haardt}, F., {Della Ceca}, R., {Volonteri}, M., \&
  {Sbarrato}, T. 2013,
  \href{http://dx.doi.org/10.1093/mnras/stt637}{\JournalTitle{\mnras}, 432,
  2818}

\bibitem[{{Gilli} {et~al.}(2019){Gilli}, {Mignoli}, {Peca}, {Nanni}, {Prand
  oni}, {Liuzzo}, {D'Amato}, {Brusa}, {Calura}, {Caminha}, {Chiaberge},
  {Comastri}, {Cucciati}, {Cusano}, {Grandi}, {Decarli}, {Lanzuisi},
  {Mannucci}, {Pinna}, {Tozzi}, {Vanzella}, {Vignali}, {Vito}, {Balmaverde},
  {Citro}, {Cappelluti}, {Zamorani}, \& {Norman}}]{2019A&A...632A..26G}
{Gilli}, R., {Mignoli}, M., {Peca}, A., {et~al.} 2019,
  \href{http://dx.doi.org/10.1051/0004-6361/201936121}{\JournalTitle{\aap},
  632, A26}

\bibitem[{{Gnedin} {et~al.}(2015){Gnedin}, {Mikhailov}, \&
  {Piotrovich}}]{2015AN....336..312G}
{Gnedin}, Y.~N., {Mikhailov}, A.~G., \& {Piotrovich}, M.~Y. 2015,
  \href{http://dx.doi.org/10.1002/asna.201412161}{\JournalTitle{Astronomische
  Nachrichten}, 336, 312}

\bibitem[{{Harris} \& {Krawczynski}(2006)}]{2006ARA&A..44..463H}
{Harris}, D.~E., \& {Krawczynski}, H. 2006,
  \href{http://dx.doi.org/10.1146/annurev.astro.44.051905.092446}{\JournalTitle{\araa},
  44, 463}

\bibitem[{{HI4PI Collaboration} {et~al.}(2016){HI4PI Collaboration}, {Ben
  Bekhti}, {Fl{\"o}er}, {Keller}, {Kerp}, {Lenz}, {Winkel}, {Bailin},
  {Calabretta}, {Dedes}, {Ford}, {Gibson}, {Haud}, {Janowiecki}, {Kalberla},
  {Lockman}, {McClure-Griffiths}, {Murphy}, {Nakanishi}, {Pisano}, \&
  {Staveley-Smith}}]{2016A&A...594A.116H}
{HI4PI Collaboration}, {Ben Bekhti}, N., {Fl{\"o}er}, L., {et~al.} 2016,
  \href{http://dx.doi.org/10.1051/0004-6361/201629178}{\JournalTitle{\aap},
  594, A116}

\bibitem[{{Hook} {et~al.}(2004){Hook}, {J{\o}rgensen}, {Allington-Smith},
  {Davies}, {Metcalfe}, {Murowinski}, \& {Crampton}}]{2004PASP..116..425H}
{Hook}, I.~M., {J{\o}rgensen}, I., {Allington-Smith}, J.~R., {et~al.} 2004,
  \href{http://dx.doi.org/10.1086/383624}{\JournalTitle{\pasp}, 116, 425}

\bibitem[{{Ighina} {et~al.}(2019){Ighina}, {Caccianiga}, {Moretti},
  {Belladitta}, {Della Ceca}, {Ballo}, \& {Dallacasa}}]{2019MNRAS.489.2732I}
{Ighina}, L., {Caccianiga}, A., {Moretti}, A., {et~al.} 2019,
  \href{http://dx.doi.org/10.1093/mnras/stz2340}{\JournalTitle{\mnras}, 489,
  2732}

\bibitem[{{Inayoshi} {et~al.}(2020){Inayoshi}, {Visbal}, \&
  {Haiman}}]{2020ARA&A..58...27I}
{Inayoshi}, K., {Visbal}, E., \& {Haiman}, Z. 2020,
  \href{http://dx.doi.org/10.1146/annurev-astro-120419-014455}{\JournalTitle{\araa},
  58, 27}

\bibitem[{{Jiang} {et~al.}(2016){Jiang}, {McGreer}, {Fan}, {Strauss},
  {Ba{\~n}ados}, {Becker}, {Bian}, {Farnsworth}, {Shen}, {Wang}, {Wang},
  {Wang}, {White}, {Wu}, {Wu}, {Yang}, \& {Yang}}]{2016ApJ...833..222J}
{Jiang}, L., {McGreer}, I.~D., {Fan}, X., {et~al.} 2016,
  \href{http://dx.doi.org/10.3847/1538-4357/833/2/222}{\JournalTitle{\apj},
  833, 222}

\bibitem[{{Jimenez-Gallardo} {et~al.}(2020){Jimenez-Gallardo}, {Massaro},
  {Prieto}, {Missaglia}, {Stuardi}, {Paggi}, {Ricci}, {Kraft}, {Liuzzo},
  {Tremblay}, {Baum}, {O'Dea}, {Wilkes}, {Kuraszkiewicz}, {Forman}, \&
  {Harris}}]{2020ApJS..250....7J}
{Jimenez-Gallardo}, A., {Massaro}, F., {Prieto}, M.~A., {et~al.} 2020,
  \href{http://dx.doi.org/10.3847/1538-4365/aba5a0}{\JournalTitle{\apjs}, 250,
  7}

\bibitem[{{Jolley} \& {Kuncic}(2008)}]{2008MNRAS.386..989J}
{Jolley}, E.~J.~D., \& {Kuncic}, Z. 2008,
  \href{http://dx.doi.org/10.1111/j.1365-2966.2008.13082.x}{\JournalTitle{\mnras},
  386, 989}

\bibitem[{{Jolley} {et~al.}(2009){Jolley}, {Kuncic}, {Bicknell}, \&
  {Wagner}}]{2009MNRAS.400.1521J}
{Jolley}, E.~J.~D., {Kuncic}, Z., {Bicknell}, G.~V., \& {Wagner}, S. 2009,
  \href{http://dx.doi.org/10.1111/j.1365-2966.2009.15554.x}{\JournalTitle{\mnras},
  400, 1521}

\bibitem[{{Kellermann} {et~al.}(1989){Kellermann}, {Sramek}, {Schmidt},
  {Shaffer}, \& {Green}}]{1989AJ.....98.1195K}
{Kellermann}, K.~I., {Sramek}, R., {Schmidt}, M., {Shaffer}, D.~B., \& {Green},
  R. 1989, \href{http://dx.doi.org/10.1086/115207}{\JournalTitle{\aj}, 98,
  1195}

\bibitem[{{Labrie} {et~al.}(2019){Labrie}, {Anderson}, {C{\'a}rdenes},
  {Simpson}, \& {Turner}}]{2019ASPC..523..321L}
{Labrie}, K., {Anderson}, K., {C{\'a}rdenes}, R., {Simpson}, C., \& {Turner},
  J. E.~H. 2019, \JournalTitle{adass XXVII}, 523, 321

\bibitem[{{Lansbury} {et~al.}(2014){Lansbury}, {Alexander}, {Del Moro}, {Gand
  hi}, {Assef}, {Stern}, {Aird}, {Ballantyne}, {Balokovi{\'c}}, {Bauer},
  {Boggs}, {Brandt}, {Christensen}, {Craig}, {Elvis}, {Grefenstette}, {Hailey},
  {Harrison}, {Hickox}, {Koss}, {LaMassa}, {Luo}, {Mullaney}, {Teng}, {Urry},
  \& {Zhang}}]{2014ApJ...785...17L}
{Lansbury}, G.~B., {Alexander}, D.~M., {Del Moro}, A., {et~al.} 2014,
  \href{http://dx.doi.org/10.1088/0004-637X/785/1/17}{\JournalTitle{\apj}, 785,
  17}

\bibitem[{{Li} {et~al.}(2003){Li}, {Kastner}, {Prigozhin}, \&
  {Schulz}}]{2003ApJ...590..586L}
{Li}, J., {Kastner}, J.~H., {Prigozhin}, G.~Y., \& {Schulz}, N.~S. 2003,
  \href{http://dx.doi.org/10.1086/374967}{\JournalTitle{\apj}, 590, 586}

\bibitem[{{Li} {et~al.}(2004){Li}, {Kastner}, {Prigozhin}, {Schulz},
  {Feigelson}, \& {Getman}}]{2004ApJ...610.1204L}
{Li}, J., {Kastner}, J.~H., {Prigozhin}, G.~Y., {et~al.} 2004,
  \href{http://dx.doi.org/10.1086/421866}{\JournalTitle{\apj}, 610, 1204}

\bibitem[{{Lucchini} {et~al.}(2017){Lucchini}, {Tavecchio}, \&
  {Ghisellini}}]{2017MNRAS.466.4299L}
{Lucchini}, M., {Tavecchio}, F., \& {Ghisellini}, G. 2017,
  \href{http://dx.doi.org/10.1093/mnras/stw3316}{\JournalTitle{\mnras}, 466,
  4299}

\bibitem[{{Lusso} \& {Risaliti}(2016)}]{2016ApJ...819..154L}
{Lusso}, E., \& {Risaliti}, G. 2016,
  \href{http://dx.doi.org/10.3847/0004-637X/819/2/154}{\JournalTitle{\apj},
  819, 154}

\bibitem[{{Marchesi} {et~al.}(2016){Marchesi}, {Civano}, {Elvis}, {Salvato},
  {Brusa}, {Comastri}, {Gilli}, {Hasinger}, {Lanzuisi}, {Miyaji}, {Treister},
  {Urry}, {Vignali}, {Zamorani}, {Allevato}, {Cappelluti}, {Cardamone},
  {Finoguenov}, {Griffiths}, {Karim}, {Laigle}, {LaMassa}, {Jahnke}, {Ranalli},
  {Schawinski}, {Schinnerer}, {Silverman}, {Smolcic}, {Suh}, \&
  {Trakhtenbrot}}]{2016ApJ...817...34M}
{Marchesi}, S., {Civano}, F., {Elvis}, M., {et~al.} 2016,
  \href{http://dx.doi.org/10.3847/0004-637X/817/1/34}{\JournalTitle{\apj}, 817,
  34}

\bibitem[{{Marshall} {et~al.}(2018){Marshall}, {Gelbord}, {Worrall},
  {Birkinshaw}, {Schwartz}, {Jauncey}, {Griffiths}, {Murphy}, {Lovell},
  {Perlman}, \& {Godfrey}}]{2018ApJ...856...66M}
{Marshall}, H.~L., {Gelbord}, J.~M., {Worrall}, D.~M., {et~al.} 2018,
  \href{http://dx.doi.org/10.3847/1538-4357/aaaf66}{\JournalTitle{\apj}, 856,
  66}

\bibitem[{{Matsuoka} {et~al.}(2019){Matsuoka}, {Iwasawa}, {Onoue}, {Kashikawa},
  {Strauss}, {Lee}, {Imanishi}, {Nagao}, {Akiyama}, {Asami}, {Bosch},
  {Furusawa}, {Goto}, {Gunn}, {Harikane}, {Ikeda}, {Izumi}, {Kawaguchi},
  {Kato}, {Kikuta}, {Kohno}, {Komiyama}, {Koyama}, {Lupton}, {Minezaki},
  {Miyazaki}, {Murayama}, {Niida}, {Nishizawa}, {Noboriguchi}, {Oguri}, {Ono},
  {Ouchi}, {Price}, {Sameshima}, {Schulze}, {Silverman}, {Sugiyama}, {Tait},
  {Takada}, {Takata}, {Tanaka}, {Tang}, {Toba}, {Utsumi}, {Wang}, \&
  {Yamashita}}]{2019ApJ...883..183M}
{Matsuoka}, Y., {Iwasawa}, K., {Onoue}, M., {et~al.} 2019,
  \href{http://dx.doi.org/10.3847/1538-4357/ab3c60}{\JournalTitle{\apj}, 883,
  183}

\bibitem[{{Mazzucchelli} {et~al.}(2017){Mazzucchelli}, {Ba{\~n}ados},
  {Venemans}, {Decarli}, {Farina}, {Walter}, {Eilers}, {Rix}, {Simcoe},
  {Stern}, {Fan}, {Schlafly}, {De Rosa}, {Hennawi}, {Chambers}, {Greiner},
  {Burgett}, {Draper}, {Kaiser}, {Kudritzki}, {Magnier}, {Metcalfe}, {Waters},
  \& {Wainscoat}}]{2017ApJ...849...91M}
{Mazzucchelli}, C., {Ba{\~n}ados}, E., {Venemans}, B.~P., {et~al.} 2017,
  \href{http://dx.doi.org/10.3847/1538-4357/aa9185}{\JournalTitle{\apj}, 849,
  91}

\bibitem[{{McKeough} {et~al.}(2016){McKeough}, {Siemiginowska}, {Cheung},
  {Stawarz}, {Kashyap}, {Stein}, {Stampoulis}, {van Dyk}, {Wardle}, {Lee},
  {Harris}, {Schwartz}, {Donato}, {Maraschi}, \&
  {Tavecchio}}]{2016ApJ...833..123M}
{McKeough}, K., {Siemiginowska}, A., {Cheung}, C.~C., {et~al.} 2016,
  \href{http://dx.doi.org/10.3847/1538-4357/833/1/123}{\JournalTitle{\apj},
  833, 123}

\bibitem[{{McLean} {et~al.}(2010){McLean}, {Steidel}, {Epps}, {Matthews},
  {Adkins}, {Konidaris}, {Weber}, {Aliado}, {Brims}, {Canfield}, {Cromer},
  {Fucik}, {Kulas}, {Mace}, {Magnone}, {Rodriguez}, {Wang}, \&
  {Weiss}}]{2010SPIE.7735E..1EM}
{McLean}, I.~S., {Steidel}, C.~C., {Epps}, H., {et~al.} 2010,
  \href{http://dx.doi.org/10.1117/12.856715}{\JournalTitle{Proc. SPIE}, 7735,
  77351E}

\bibitem[{{McLean} {et~al.}(2012){McLean}, {Steidel}, {Epps}, {Konidaris},
  {Matthews}, {Adkins}, {Aliado}, {Brims}, {Canfield}, {Cromer}, {Fucik},
  {Kulas}, {Mace}, {Magnone}, {Rodriguez}, {Rudie}, {Trainor}, {Wang}, {Weber},
  \& {Weiss}}]{2012SPIE.8446E..0JM}
{McLean}, I.~S., {Steidel}, C.~C., {Epps}, H.~W., {et~al.} 2012,
  \href{http://dx.doi.org/10.1117/12.924794}{\JournalTitle{Proc. SPIE}, 8446,
  84460J}

\bibitem[{{Medvedev} {et~al.}(2020){Medvedev}, {Sazonov}, {Gilfanov},
  {Burenin}, {Khorunzhev}, {Meshcheryakov}, {Sunyaev}, {Bikmaev}, \&
  {Irtuganov}}]{2020MNRAS.497.1842M}
{Medvedev}, P., {Sazonov}, S., {Gilfanov}, M., {et~al.} 2020,
  \href{http://dx.doi.org/10.1093/mnras/staa2051}{\JournalTitle{\mnras}, 497,
  1842}

\bibitem[{{Meyer} {et~al.}(2015){Meyer}, {Georganopoulos}, {Sparks}, {Godfrey},
  {Lovell}, \& {Perlman}}]{2015ApJ...805..154M}
{Meyer}, E.~T., {Georganopoulos}, M., {Sparks}, W.~B., {et~al.} 2015,
  \href{http://dx.doi.org/10.1088/0004-637X/805/2/154}{\JournalTitle{\apj},
  805, 154}

\bibitem[{{Miley}(1980)}]{1980ARA&A..18..165M}
{Miley}, G. 1980,
  \href{http://dx.doi.org/10.1146/annurev.aa.18.090180.001121}{\JournalTitle{\araa},
  18, 165}

\bibitem[{{Miller} {et~al.}(2011){Miller}, {Brandt}, {Schneider}, {Gibson},
  {Steffen}, \& {Wu}}]{2011ApJ...726...20M}
{Miller}, B.~P., {Brandt}, W.~N., {Schneider}, D.~P., {et~al.} 2011,
  \href{http://dx.doi.org/10.1088/0004-637X/726/1/20}{\JournalTitle{\apj}, 726,
  20}

\bibitem[{{Momjian} {et~al.}(2018){Momjian}, {Carilli}, {Ba{\~n}ados},
  {Walter}, \& {Venemans}}]{2018ApJ...861...86M}
{Momjian}, E., {Carilli}, C.~L., {Ba{\~n}ados}, E., {Walter}, F., \&
  {Venemans}, B.~P. 2018,
  \href{http://dx.doi.org/10.3847/1538-4357/aac76f}{\JournalTitle{\apj}, 861,
  86}

\bibitem[{{Mortlock} {et~al.}(2011){Mortlock}, {Warren}, {Venemans}, {Patel},
  {Hewett}, {McMahon}, {Simpson}, {Theuns}, {Gonz{\'a}les-Solares}, {Adamson},
  {Dye}, {Hambly}, {Hirst}, {Irwin}, {Kuiper}, {Lawrence}, \&
  {R{\"o}ttgering}}]{2011Natur.474..616M}
{Mortlock}, D.~J., {Warren}, S.~J., {Venemans}, B.~P., {et~al.} 2011,
  \href{http://dx.doi.org/10.1038/nature10159}{\JournalTitle{\nat}, 474, 616}

\bibitem[{{Mushotzky} {et~al.}(2019){Mushotzky}, {Aird}, {Barger},
  {Cappelluti}, {Chartas}, {Corrales}, {Eufrasio}, {Fabian}, {Falcone},
  {Gallo}, {Gilli}, {Grant}, {Hardcastle}, {Hodges-Kluck}, {Kara}, {Koss},
  {Li}, {Lisse}, {Loewenstein}, {Markevitch}, {Meyer}, {Miller}, {Mulchaey},
  {Petre}, {Ptak}, {Reynolds}, {Russell}, {Safi-Harb}, {Smith}, {Snios},
  {Tombesi}, {Valencic}, {Walker}, {Williams}, {Winter}, {Yamaguchi}, {Zhang},
  {Arenberg}, {Brand t}, {Burrows}, {Georganopoulos}, {Miller}, {Norman}, \&
  {Rosati}}]{2019BAAS...51g.107M}
{Mushotzky}, R., {Aird}, J., {Barger}, A.~J., {et~al.} 2019, in \baas, Vol.~51,
  107

\bibitem[{{Nandra} {et~al.}(2013){Nandra}, {Barret}, {Barcons}, {Fabian}, {den
  Herder}, {Piro}, {Watson}, {Adami}, {Aird}, {Afonso}, {Alexander},
  {Argiroffi}, {Amati}, {Arnaud}, {Atteia}, {Audard}, {Badenes}, {Ballet},
  {Ballo}, {Bamba}, {Bhardwaj}, {Stefano Battistelli}, {Becker}, {De Becker},
  {Behar}, {Bianchi}, {Biffi}, {B{\^\i}rzan}, {Bocchino}, {Bogdanov}, {Boirin},
  {Boller}, {Borgani}, {Borm}, {Bouch{\'e}}, {Bourdin}, {Bower}, {Braito},
  {Branchini}, {Branduardi-Raymont}, {Bregman}, {Brenneman}, {Brightman},
  {Br{\"u}ggen}, {Buchner}, {Bulbul}, {Brusa}, {Bursa}, {Caccianiga},
  {Cackett}, {Campana}, {Cappelluti}, {Cappi}, {Carrera}, {Ceballos},
  {Christensen}, {Chu}, {Churazov}, {Clerc}, {Corbel}, {Corral}, {Comastri},
  {Costantini}, {Croston}, {Dadina}, {D'Ai}, {Decourchelle}, {Della Ceca},
  {Dennerl}, {Dolag}, {Done}, {Dovciak}, {Drake}, {Eckert}, {Edge}, {Ettori},
  {Ezoe}, {Feigelson}, {Fender}, {Feruglio}, {Finoguenov}, {Fiore}, {Galeazzi},
  {Gallagher}, {Gandhi}, {Gaspari}, {Gastaldello}, {Georgakakis},
  {Georgantopoulos}, {Gilfanov}, {Gitti}, {Gladstone}, {Goosmann}, {Gosset},
  {Grosso}, {Guedel}, {Guerrero}, {Haberl}, {Hardcastle}, {Heinz}, {Alonso
  Herrero}, {Herv{\'e}}, {Holmstrom}, {Iwasawa}, {Jonker}, {Kaastra}, {Kara},
  {Karas}, {Kastner}, {King}, {Kosenko}, {Koutroumpa}, {Kraft}, {Kreykenbohm},
  {Lallement}, {Lanzuisi}, {Lee}, {Lemoine-Goumard}, {Lobban}, {Lodato},
  {Lovisari}, {Lotti}, {McCharthy}, {McNamara}, {Maggio}, {Maiolino}, {De
  Marco}, {de Martino}, {Mateos}, {Matt}, {Maughan}, {Mazzotta}, {Mendez},
  {Merloni}, {Micela}, {Miceli}, {Mignani}, {Miller}, {Miniutti}, {Molendi},
  {Montez}, {Moretti}, {Motch}, {Naz{\'e}}, {Nevalainen}, {Nicastro}, {Nulsen},
  {Ohashi}, {O'Brien}, {Osborne}, {Oskinova}, {Pacaud}, {Paerels}, {Page},
  {Papadakis}, {Pareschi}, {Petre}, {Petrucci}, {Piconcelli}, {Pillitteri},
  {Pinto}, {de Plaa}, {Pointecouteau}, {Ponman}, {Ponti}, {Porquet}, {Pounds},
  {Pratt}, {Predehl}, {Proga}, {Psaltis}, {Rafferty}, {Ramos-Ceja}, {Ranalli},
  {Rasia}, {Rau}, {Rauw}, {Rea}, {Read}, {Reeves}, {Reiprich}, {Renaud},
  {Reynolds}, {Risaliti}, {Rodriguez}, {Rodriguez Hidalgo}, {Roncarelli},
  {Rosario}, {Rossetti}, {Rozanska}, {Rovilos}, {Salvaterra}, {Salvato}, {Di
  Salvo}, {Sanders}, {Sanz-Forcada}, {Schawinski}, {Schaye}, {Schwope},
  {Sciortino}, {Severgnini}, {Shankar}, {Sijacki}, {Sim}, {Schmid}, {Smith},
  {Steiner}, {Stelzer}, {Stewart}, {Strohmayer}, {Str{\"u}der}, {Sun}, {Takei},
  {Tatischeff}, {Tiengo}, {Tombesi}, {Trinchieri}, {Tsuru}, {Ud-Doula},
  {Ursino}, {Valencic}, {Vanzella}, {Vaughan}, {Vignali}, {Vink}, {Vito},
  {Volonteri}, {Wang}, {Webb}, {Willingale}, {Wilms}, {Wise}, {Worrall},
  {Young}, {Zampieri}, {In't Zand}, {Zane}, {Zezas}, {Zhang}, \&
  {Zhuravleva}}]{2013arXiv1306.2307N}
{Nandra}, K., {Barret}, D., {Barcons}, X., {et~al.} 2013, \JournalTitle{arXiv
  e-prints}, arXiv:1306.2307

\bibitem[{{Nanni} {et~al.}(2018){Nanni}, {Gilli}, {Vignali}, {Mignoli},
  {Comastri}, {Vanzella}, {Zamorani}, {Calura}, {Lanzuisi}, {Brusa}, {Tozzi},
  {Iwasawa}, {Cappi}, {Vito}, {Balmaverde}, {Costa}, {Risaliti}, {Paolillo},
  {Prandoni}, {Liuzzo}, {Rosati}, {Chiaberge}, {Caminha}, {Sani}, {Cappelluti},
  \& {Norman}}]{2018A&A...614A.121N}
{Nanni}, R., {Gilli}, R., {Vignali}, C., {et~al.} 2018,
  \href{http://dx.doi.org/10.1051/0004-6361/201832694}{\JournalTitle{\aap},
  614, A121}

\bibitem[{{Napier} {et~al.}(2020){Napier}, {Foord}, {Gallo}, {Ghisellini},
  {Hodges-Kluck}, {Wu}, {Haardt}, \& {Ciardi}}]{2020MNRAS.498.1550N}
{Napier}, K., {Foord}, A., {Gallo}, E., {et~al.} 2020,
  \href{http://dx.doi.org/10.1093/mnras/staa2178}{\JournalTitle{\mnras}, 498,
  1550}

\bibitem[{{Nesvadba} {et~al.}(2020){Nesvadba}, {Bicknell}, {Mukherjee}, \&
  {Wagner}}]{2020A&A...639L..13N}
{Nesvadba}, N.~P.~H., {Bicknell}, G.~V., {Mukherjee}, D., \& {Wagner}, A.~Y.
  2020,
  \href{http://dx.doi.org/10.1051/0004-6361/202038269}{\JournalTitle{\aap},
  639, L13}

\bibitem[{{Paliya} {et~al.}(2020){Paliya}, {Ajello}, {Cao}, {Giroletti},
  {Kaur}, {Madejski}, {Lott}, \& {Hartmann}}]{2020ApJ...897..177P}
{Paliya}, V.~S., {Ajello}, M., {Cao}, H.~M., {et~al.} 2020,
  \href{http://dx.doi.org/10.3847/1538-4357/ab9c1a}{\JournalTitle{\apj}, 897,
  177}

\bibitem[{{Park} {et~al.}(2006){Park}, {Kashyap}, {Siemiginowska}, {van Dyk},
  {Zezas}, {Heinke}, \& {Wargelin}}]{2006ApJ...652..610P}
{Park}, T., {Kashyap}, V.~L., {Siemiginowska}, A., {et~al.} 2006,
  \href{http://dx.doi.org/10.1086/507406}{\JournalTitle{\apj}, 652, 610}

\bibitem[{{Pons} {et~al.}(2019){Pons}, {McMahon}, {Simcoe}, {Banerji},
  {Hewett}, \& {Reed}}]{2019MNRAS.484.5142P}
{Pons}, E., {McMahon}, R.~G., {Simcoe}, R.~A., {et~al.} 2019,
  \href{http://dx.doi.org/10.1093/mnras/stz292}{\JournalTitle{\mnras}, 484,
  5142}

\bibitem[{{Reddy} {et~al.}(2021){Reddy}, {Georganopoulos}, \&
  {Meyer}}]{2021arXiv210102024R}
{Reddy}, K., {Georganopoulos}, M., \& {Meyer}, E.~T. 2021, \JournalTitle{arXiv
  e-prints}, arXiv:2101.02024

\bibitem[{{Reed} {et~al.}(2017){Reed}, {McMahon}, {Martini}, {Banerji},
  {Auger}, {Hewett}, {Koposov}, {Gibbons}, {Gonzalez-Solares}, {Ostrovski},
  {Tie}, {Abdalla}, {Allam}, {Benoit-L{\'e}vy}, {Bertin}, {Brooks},
  {Buckley-Geer}, {Burke}, {Carnero Rosell}, {Carrasco Kind}, {Carretero}, {da
  Costa}, {DePoy}, {Desai}, {Diehl}, {Doel}, {Evrard}, {Finley}, {Flaugher},
  {Fosalba}, {Frieman}, {Garc{\'\i}a-Bellido}, {Gaztanaga}, {Goldstein},
  {Gruen}, {Gruendl}, {Gutierrez}, {James}, {Kuehn}, {Kuropatkin}, {Lahav},
  {Lima}, {Maia}, {Marshall}, {Melchior}, {Miller}, {Miquel}, {Nord}, {Ogando},
  {Plazas}, {Romer}, {Sanchez}, {Scarpine}, {Schubnell}, {Sevilla-Noarbe},
  {Smith}, {Sobreira}, {Suchyta}, {Swanson}, {Tarle}, {Tucker}, {Walker}, \&
  {Wester}}]{2017MNRAS.468.4702R}
{Reed}, S.~L., {McMahon}, R.~G., {Martini}, P., {et~al.} 2017,
  \href{http://dx.doi.org/10.1093/mnras/stx728}{\JournalTitle{\mnras}, 468,
  4702}

\bibitem[{{Regan} {et~al.}(2019){Regan}, {Downes}, {Volonteri}, {Beckmann},
  {Lupi}, {Trebitsch}, \& {Dubois}}]{2019MNRAS.486.3892R}
{Regan}, J.~A., {Downes}, T.~P., {Volonteri}, M., {et~al.} 2019,
  \href{http://dx.doi.org/10.1093/mnras/stz1045}{\JournalTitle{\mnras}, 486,
  3892}

\bibitem[{{Schwartz} {et~al.}(2000){Schwartz}, {Marshall}, {Lovell}, {Piner},
  {Tingay}, {Birkinshaw}, {Chartas}, {Elvis}, {Feigelson}, {Ghosh}, {Harris},
  {Hirabayashi}, {Hooper}, {Jauncey}, {Lanzetta}, {Mathur}, {Preston},
  {Tucker}, {Virani}, {Wilkes}, \& {Worrall}}]{2000ApJ...540L..69S}
{Schwartz}, D.~A., {Marshall}, H.~L., {Lovell}, J.~E.~J., {et~al.} 2000,
  \href{http://dx.doi.org/10.1086/312875}{\JournalTitle{\apjl}, 540, 69}

\bibitem[{{Schwartz} {et~al.}(2020){Schwartz}, {Siemiginowska}, {Snios},
  {Worrall}, {Birkinshaw}, {Cheung}, {Marshall}, {Migliori}, {Wardle}, \&
  {Gobeille}}]{2020ApJ...904...57S}
{Schwartz}, D.~A., {Siemiginowska}, A., {Snios}, B., {et~al.} 2020,
  \href{http://dx.doi.org/10.3847/1538-4357/abbd99}{\JournalTitle{\apj}, 904,
  57}

\bibitem[{{Siemiginowska} {et~al.}(2003){Siemiginowska}, {Smith}, {Aldcroft},
  {Schwartz}, {Paerels}, \& {Petric}}]{2003ApJ...598L..15S}
{Siemiginowska}, A., {Smith}, R.~K., {Aldcroft}, T.~L., {et~al.} 2003,
  \href{http://dx.doi.org/10.1086/380497}{\JournalTitle{\apjl}, 598, L15}

\bibitem[{{Siemiginowska} {et~al.}(2007){Siemiginowska}, {Stawarz}, {Cheung},
  {Harris}, {Sikora}, {Aldcroft}, \& {Bechtold}}]{2007ApJ...657..145S}
{Siemiginowska}, A., {Stawarz}, {\L}., {Cheung}, C.~C., {et~al.} 2007,
  \href{http://dx.doi.org/10.1086/510898}{\JournalTitle{\apj}, 657, 145}

\bibitem[{{Sikora} {et~al.}(2013){Sikora}, {Stasi{\'n}ska},
  {Kozie{\l}-Wierzbowska}, {Madejski}, \& {Asari}}]{2013ApJ...765...62S}
{Sikora}, M., {Stasi{\'n}ska}, G., {Kozie{\l}-Wierzbowska}, D., {Madejski},
  G.~M., \& {Asari}, N.~V. 2013,
  \href{http://dx.doi.org/10.1088/0004-637X/765/1/62}{\JournalTitle{\apj}, 765,
  62}

\bibitem[{{Sikora} {et~al.}(2007){Sikora}, {Stawarz}, \&
  {Lasota}}]{2007ApJ...658..815S}
{Sikora}, M., {Stawarz}, {\L}., \& {Lasota}, J.-P. 2007,
  \href{http://dx.doi.org/10.1086/511972}{\JournalTitle{\apj}, 658, 815}

\bibitem[{{Simionescu} {et~al.}(2016){Simionescu}, {Stawarz}, {Ichinohe},
  {Cheung}, {Jamrozy}, {Siemiginowska}, {Hagino}, {Gandhi}, \&
  {Werner}}]{2016ApJ...816L..15S}
{Simionescu}, A., {Stawarz}, {\L}., {Ichinohe}, Y., {et~al.} 2016,
  \href{http://dx.doi.org/10.3847/2041-8205/816/1/L15}{\JournalTitle{\apjl},
  816, L15}

\bibitem[{{Skrutskie} {et~al.}(2006){Skrutskie}, {Cutri}, {Stiening},
  {Weinberg}, {Schneider}, {Carpenter}, {Beichman}, {Capps}, {Chester},
  {Elias}, {Huchra}, {Liebert}, {Lonsdale}, {Monet}, {Price}, {Seitzer},
  {Jarrett}, {Kirkpatrick}, {Gizis}, {Howard}, {Evans}, {Fowler}, {Fullmer},
  {Hurt}, {Light}, {Kopan}, {Marsh}, {McCallon}, {Tam}, {Van Dyk}, \&
  {Wheelock}}]{2006AJ....131.1163S}
{Skrutskie}, M.~F., {Cutri}, R.~M., {Stiening}, R., {et~al.} 2006,
  \href{http://dx.doi.org/10.1086/498708}{\JournalTitle{\aj}, 131, 1163}

\bibitem[{{Spingola} {et~al.}(2020){Spingola}, {Dallacasa}, {Belladitta},
  {Caccianiga}, {Giroletti}, {Moretti}, \& {Orienti}}]{2020A&A...643L..12S}
{Spingola}, C., {Dallacasa}, D., {Belladitta}, S., {et~al.} 2020,
  \href{http://dx.doi.org/10.1051/0004-6361/202039458}{\JournalTitle{\aap},
  643, L12}

\bibitem[{{Steffen} {et~al.}(2006){Steffen}, {Strateva}, {Brandt}, {Alexander},
  {Koekemoer}, {Lehmer}, {Schneider}, \& {Vignali}}]{2006AJ....131.2826S}
{Steffen}, A.~T., {Strateva}, I., {Brandt}, W.~N., {et~al.} 2006,
  \href{http://dx.doi.org/10.1086/503627}{\JournalTitle{\aj}, 131, 2826}

\bibitem[{{Stern} {et~al.}(2000){Stern}, {Djorgovski}, {Perley}, {de Carvalho},
  \& {Wall}}]{2000AJ....119.1526S}
{Stern}, D., {Djorgovski}, S.~G., {Perley}, R.~A., {de Carvalho}, R.~R., \&
  {Wall}, J.~V. 2000,
  \href{http://dx.doi.org/10.1086/301316}{\JournalTitle{\aj}, 119, 1526}

\bibitem[{{Strateva} {et~al.}(2005){Strateva}, {Brandt}, {Schneider}, {Vanden
  Berk}, \& {Vignali}}]{2005AJ....130..387S}
{Strateva}, I.~V., {Brandt}, W.~N., {Schneider}, D.~P., {Vanden Berk}, D.~G.,
  \& {Vignali}, C. 2005,
  \href{http://dx.doi.org/10.1086/431247}{\JournalTitle{\aj}, 130, 387}

\bibitem[{{STScI development Team}(2018)}]{2018ascl.soft11001S}
{STScI development Team}. 2018, {synphot: Synthetic photometry using Astropy}

\bibitem[{{Takeo} {et~al.}(2020){Takeo}, {Inayoshi}, \&
  {Mineshige}}]{2020MNRAS.497..302T}
{Takeo}, E., {Inayoshi}, K., \& {Mineshige}, S. 2020,
  \href{http://dx.doi.org/10.1093/mnras/staa1906}{\JournalTitle{\mnras}, 497,
  302}

\bibitem[{{Tchekhovskoy} \& {McKinney}(2012)}]{2012MNRAS.423L..55T}
{Tchekhovskoy}, A., \& {McKinney}, J.~C. 2012,
  \href{http://dx.doi.org/10.1111/j.1745-3933.2012.01256.x}{\JournalTitle{\mnras},
  423, L55}

\bibitem[{{van Breugel} {et~al.}(1985){van Breugel}, {Filippenko}, {Heckman},
  \& {Miley}}]{1985ApJ...293...83V}
{van Breugel}, W., {Filippenko}, A.~V., {Heckman}, T., \& {Miley}, G. 1985,
  \href{http://dx.doi.org/10.1086/163216}{\JournalTitle{\apj}, 293, 83}

\bibitem[{{Vignali} {et~al.}(2003){Vignali}, {Brandt}, \&
  {Schneider}}]{2003AJ....125..433V}
{Vignali}, C., {Brandt}, W.~N., \& {Schneider}, D.~P. 2003,
  \href{http://dx.doi.org/10.1086/345973}{\JournalTitle{\aj}, 125, 433}

\bibitem[{{Vito} {et~al.}(2019{\natexlab{a}}){Vito}, {Brandt}, {Bauer},
  {Gilli}, {Luo}, {Zamorani}, {Calura}, {Comastri}, {Mazzucchelli}, {Mignoli},
  {Nanni}, {Shemmer}, {Vignali}, {Brusa}, {Cappelluti}, {Civano}, \&
  {Volonteri}}]{2019A&A...628L...6V}
{Vito}, F., {Brandt}, W.~N., {Bauer}, F.~E., {et~al.} 2019{\natexlab{a}},
  \href{http://dx.doi.org/10.1051/0004-6361/201935924}{\JournalTitle{\aap},
  628, L6}

\bibitem[{{Vito} {et~al.}(2019{\natexlab{b}}){Vito}, {Brandt}, {Bauer},
  {Calura}, {Gilli}, {Luo}, {Shemmer}, {Vignali}, {Zamorani}, {Brusa},
  {Civano}, {Comastri}, \& {Nanni}}]{2019A&A...630A.118V}
{Vito}, F., {Brandt}, W.~N., {Bauer}, F.~E., {et~al.} 2019{\natexlab{b}},
  \href{http://dx.doi.org/10.1051/0004-6361/201936217}{\JournalTitle{\aap},
  630, A118}

\bibitem[{{Wachter} {et~al.}(1979){Wachter}, {Leach}, \&
  {Kellogg}}]{1979ApJ...230..274W}
{Wachter}, K., {Leach}, R., \& {Kellogg}, E. 1979,
  \href{http://dx.doi.org/10.1086/157084}{\JournalTitle{\apj}, 230, 274}

\bibitem[{{Wang} {et~al.}(2019){Wang}, {Yang}, {Fan}, {Wu}, {Yue}, {Li},
  {Bian}, {Jiang}, {Ba{\~n}ados}, {Schindler}, {Findlay}, {Davies}, {Decarli},
  {Farina}, {Green}, {Hennawi}, {Huang}, {Mazzuccheli}, {McGreer}, {Venemans},
  {Walter}, {Dye}, {Lyke}, {Myers}, \& {Haze Nunez}}]{2019ApJ...884...30W}
{Wang}, F., {Yang}, J., {Fan}, X., {et~al.} 2019,
  \href{http://dx.doi.org/10.3847/1538-4357/ab2be5}{\JournalTitle{\apj}, 884,
  30}

\bibitem[{{Wang} {et~al.}(2021{\natexlab{a}}){Wang}, {Yang}, {Fan}, {Hennawi},
  {Barth}, {Banados}, {Bian}, {Boutsia}, {Connor}, {Davies}, {Decarli},
  {Eilers}, {Farina}, {Green}, {Jiang}, {Li}, {Mazzucchelli}, {Nanni},
  {Schindler}, {Venemans}, {Walter}, {Wu}, \& {Yue}}]{2021ApJ...907L...1W}
{Wang}, F., {Yang}, J., {Fan}, X., {et~al.} 2021{\natexlab{a}},
  \href{http://dx.doi.org/10.3847/2041-8213/abd8c6}{\JournalTitle{\apjl}, 907,
  L1}

\bibitem[{{Wang} {et~al.}(2021{\natexlab{b}}){Wang}, {Fan}, {Yang},
  {Mazzucchelli}, {Wu}, {Li}, {Ba{\~n}ados}, {Farina}, {Nanni}, {Ai}, {Bian},
  {Davies}, {Decarli}, {Hennawi}, {Schindler}, {Venemans}, \&
  {Walter}}]{2021ApJ...908...53W}
{Wang}, F., {Fan}, X., {Yang}, J., {et~al.} 2021{\natexlab{b}},
  \href{http://dx.doi.org/10.3847/1538-4357/abcc5e}{\JournalTitle{\apj}, 908,
  53}

\bibitem[{{Weisskopf} {et~al.}(2007){Weisskopf}, {Wu}, {Trimble}, {O'Dell},
  {Elsner}, {Zavlin}, \& {Kouveliotou}}]{2007ApJ...657.1026W}
{Weisskopf}, M.~C., {Wu}, K., {Trimble}, V., {et~al.} 2007,
  \href{http://dx.doi.org/10.1086/510776}{\JournalTitle{\apj}, 657, 1026}

\bibitem[{{Worrall} {et~al.}(2020){Worrall}, {Birkinshaw}, {Marshall},
  {Schwartz}, {Siemiginowska}, \& {Wardle}}]{2020MNRAS.497..988W}
{Worrall}, D.~M., {Birkinshaw}, M., {Marshall}, H.~L., {et~al.} 2020,
  \href{http://dx.doi.org/10.1093/mnras/staa1975}{\JournalTitle{\mnras}, 497,
  988}

\bibitem[{{Wu} {et~al.}(2015){Wu}, {Wang}, {Fan}, {Yi}, {Zuo}, {Bian}, {Jiang},
  {McGreer}, {Wang}, {Yang}, {Yang}, {Thompson}, \&
  {Beletsky}}]{2015Natur.518..512W}
{Wu}, X.-B., {Wang}, F., {Fan}, X., {et~al.} 2015,
  \href{http://dx.doi.org/10.1038/nature14241}{\JournalTitle{\nat}, 518, 512}

\bibitem[{{Yang} {et~al.}(2019){Yang}, {Wang}, {Fan}, {Yue}, {Wu}, {Li},
  {Bian}, {Jiang}, {Ba{\~n}ados}, \& {Beletsky}}]{2019AJ....157..236Y}
{Yang}, J., {Wang}, F., {Fan}, X., {et~al.} 2019,
  \href{http://dx.doi.org/10.3847/1538-3881/ab1be1}{\JournalTitle{\aj}, 157,
  236}

\bibitem[{{Yang} {et~al.}(2020){Yang}, {Wang}, {Fan}, {Hennawi}, {Davies},
  {Yue}, {Banados}, {Wu}, {Venemans}, {Barth}, {Bian}, {Boutsia}, {Decarli},
  {Farina}, {Green}, {Jiang}, {Li}, {Mazzucchelli}, \&
  {Walter}}]{2020ApJ...897L..14Y}
{Yang}, J., {Wang}, F., {Fan}, X., {et~al.} 2020,
  \href{http://dx.doi.org/10.3847/2041-8213/ab9c26}{\JournalTitle{\apjl}, 897,
  L14}

\end{thebibliography}
\end{document}